\documentclass[12pt]{iopart}
\usepackage{iopams}
\usepackage{epsfig}
\usepackage{bm}
\usepackage{psfrag}
\usepackage{multirow}
\usepackage{color}
\begin{document}

\title[Selective sweeps in growing microbial colonies]{Selective sweeps in growing microbial colonies}

\author{Kirill S Korolev$^{1,2}$\footnote{Present address: Department of Physics, Massachusetts Institute of Technology, Cambridge, Massachusetts 02139, USA}\footnotemark, Melanie J I M\"uller$^{1,2,3}$\footnote[3]{These authors contributed equally to this work.}, Nilay Karahan$^{3}$, Andrew W Murray$^{1,3}$, Oskar Hallatschek$^{4}$ and David R Nelson$^{1,2,3}$}
\address{$^{1}$ FAS Center for Systems Biology, Harvard University, Cambridge, Massachusetts 02138, USA}
\address{$^{2}$ Department of Physics, Harvard University, Cambridge, Massachusetts 02138, USA}
\address{$^{3}$ Department of Molecular and Cellular Biology, Harvard University, Cambridge, Massachusetts 02138, USA}
\address{$^{4}$ Max Planck Research Group for Biological Physics and Evolutionary Dynamics, Max Planck Institute for Dynamics \& Self-Organization (MPIDS), G\"ottingen, Germany}
\eads{\mailto{papers.korolev@gmail.com (K S Korolev)}, \mailto{mmueller@physics.harvard.edu (M J I M{\"u}ller)}, \mailto{nelson@physics.harvard.edu (D R Nelson)}}

\begin{abstract}
Evolutionary experiments with microbes are a powerful tool to study mutations and natural selection. These experiments, however, are often limited to the well-mixed environments of a test tube or a chemostat. Since spatial organization can significantly affect evolutionary dynamics, the need is growing for evolutionary experiments in spatially structured environments. The surface of a Petri dish provides such an environment, but a more detailed understanding of microbial growth on Petri dishes is necessary to interpret such experiments. We formulate a simple deterministic reaction-diffusion model, which successfully predicts the spatial patterns created by two competing species during colony expansion. We also derive the shape of these patterns analytically without relying on microscopic details of the model. In particular, we find that the relative fitness of two microbial strains can be estimated from the logarithmic spirals created by selective sweeps. The theory is tested with strains of the budding yeast \textit{Saccharomyces cerevisiae}, for spatial competitions with different initial conditions and for a range of relative fitnesses. The reaction-diffusion model also connects the microscopic parameters like growth rates and diffusion constants with macroscopic spatial patterns and predicts the relationship between fitness in liquid cultures and on Petri dishes, which we confirmed experimentally. Spatial sector patterns therefore provide an alternative fitness assay to the commonly used liquid culture fitness assays.
\end{abstract}

\pacs{87.23.Kg, 87.23.Cc, 87.18.Hf, 87.18.Tt}
\vspace{2pc}
\noindent{\it Keywords: Fisher waves, wave velocity, selective sweep, competition at the front, spatial assay, relative fitness}
\maketitle
\section{Introduction}
\label{SIntroduction}

Traditionally, the theory of evolution has been developed by analyzing phenotypes and genotypes found in natural populations and fossil records~\cite{Darwin:Origin,Barton:Evolution}. However, due to recent developments in microbiology and modern genetics, evolutionary experiments are becoming a valuable research tool~\cite{Elena:EvolutionReview}. Laboratory experiments hold great promise for uncovering basic evolutionary mechanisms by allowing us to observe evolution over time. More important, experiments, unlike evolution in natural populations, can be repeated systematically to distinguish between general principles and historical accidents. Microbes are particularly suited for evolutionary studies because they are relatively simple, reproduce and evolve rapidly, and can be easily modified using genetic engineering. Experiments with microorganisms could also provide insights into tumor growth, the spread of antibiotic resistance, and directed evolution of microbes to produce medicines or biofuels~\cite{Ron:Sectors,lambert:analogy}. 

One potential drawback of evolutionary experiments is that they are conducted in artificial laboratory environments, which are quite different from the natural ecology of the species studied. The choice of the laboratory environment is therefore very important because it could affect both the nature of observed adaptations and the evolutionary dynamics. Most microbial experiments, of interest to us here, are conducted in the well-mixed environments of a chemostat or a test tube. These well-controlled environments allow researchers to compare experimental results to theoretical predictions, but it is important to ensure that such results are generic, not environment specific. Spatial structure, absent in well-mixed cultures, can significantly affect evolutionary dynamics~\cite{Korolev:Review,kerr:tragedy,Coberly:SpaceCoEvolution,hauert:snowdrift,rainey:radiation}; therefore, it is important to carry out experiments with growth conditions that allow spatial inhomogeneities to form. The surface of a Petri dish is an easy-to-use environment, chemically similar to the environments of a test tube or a chemostat, yet capable of sustaining and preserving spatial structure during colony growth. Recently, several studies have used microorganisms in Petri dishes to study spatial patterning, mutations, and evolution~\cite{wakita:expansion,benjacob:cooperative_growth,benjacob:complex_patterns,Ron:Sectors,golding:sectors,HallatschekNelson:ExperimentalSegregation,Coberly:SpaceCoEvolution,beer:deadly,korolev:neutral_expansions,KerrBohannan02,HabetsDeVisser06,Harcombe10,SaxerTravisano09,VelicerYu03}. Our focus here is on \textit{compact} growth of colonies containing different genetic variants. For a comprehensive review emphasizing the beautiful dendritic growth patterns that can arise at low nutrient concentrations see Ref.~\cite{benjacob:review}.

An important obstacle to a wider use of spatial environments in evolutionary experiments is the limited theoretical understanding of how basic evolutionary processes play out in a spatial context. In particular, one must have a way to measure fitness to study evolution, but there are few models that relate microscopic parameters of the organisms to macroscopic quantities that can be easily measured in the laboratory. In this paper, we formulate a coarse-grained model of spatial competition that fills this gap and carry out microbial experiments to test the model's predictions. This model is based on deterministic reaction-diffusion equations, which describe short-range migration of the organisms and their competition. Our numerical results are further supported by a geometric argument, which does not rely on the detailed assumptions about microbial growth and migration. Stochastic effects due to number fluctuations~(genetic drift) can be included in the model~\cite{HallatschekNelson:ExperimentalSegregation,Hallatschek:LifeFront,hallatschek:noisy_wave}, but they do not play a central role in our analysis and are neglected for simplicity. For a study of spatial competition between two neutral bacterial strains, which is dominated by genetic drift, see Ref.~\cite{korolev:neutral_expansions}. Although we focus on competitions, the model is sufficiently general to describe mutualistic and antagonistic interactions as well.

Compared to well-mixed populations, spatial populations have a wide range of initial conditions because one has to specify not only the relative fractions of genotypes but also their spatial distribution at time zero. We examine several experimentally interesting initial conditions and calculate how the spatial distributions of genotypes changes with time. We find that the shape of the resulting spatio-genetic patterns is determined by the expansion velocities ratio~$v_{1}/v_{2}$, where~$v_{1}$ and~$v_{2}$ are the expansion velocities of isolated colonies composed exclusively of strain~$1$ or strain~$2$. Relative fitnesses of the genotypes can then be estimated by comparing experimentally observed spatio-genetic patterns to the theory. In particular, the selective advantage in this context can be defined as~$s=v_{1}/v_{2}-1$. As we show below, this definition is closely related to the traditional definition in terms of exponential growth rates. For linear inoculations, our results in the long time limit agree with Ref.~\cite{Hallatschek:LifeFront}, where a phenomenological model of the patterns was first proposed. In contrast to Ref.~\cite{Hallatschek:LifeFront}, we also study circular inoculations and carry out experiments to test quantitative predictions of our model.

\begin{figure}
\includegraphics[width=8cm]{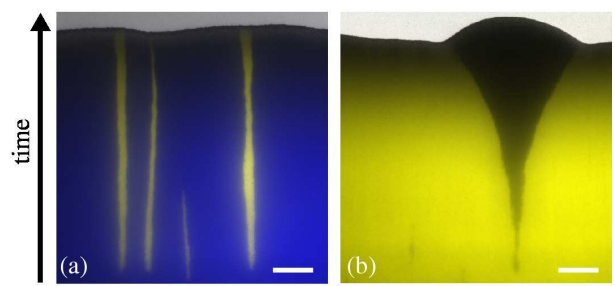}
\caption{(Colour online) Comparison of spatial segregation during a range expansion of Baker's yeast,~\textit{Saccharomyces cerevisiae} with (a)~equal and (b)~different growth rates of the two competing strains. The Petri dishes were inoculated with a well-mixed population occupying a narrow horizontal linear region at the bottom of the images from which the sectors appear. As the populations expand, they segregate into well defined domains. Different colours label different genotypes. In (a), the two strains~(yellow and blue) have the same fitness and the demixing is driven primarily by number fluctuations (genetic drift)~\protect{\cite{HallatschekNelson:ExperimentalSegregation,Korolev:Review}}. It is likely that the small variations with horizontal position in boundary slopes are related to undulations of linear fronts, which are hard to suppress when the front is very long~\cite{Kessler:Instabilities}. In (b), the sector is formed by the fitter strain~(black), and the sector expansion is caused by the difference in growth rates of the strains, or, in other words, by natural selection. In both~(a) and~(b), the scale bars are 500$\;\mu$m.}
\label{FNSComparison}
\end{figure}

The theory can be directly compared to the experiments because the spatial distribution of genotypes on a Petri dish can be visualized and quantified with fluorescent markers~\cite{HallatschekNelson:ExperimentalSegregation}. Here, we briefly describe this technique; see the supplementary information (section~S1) for more details. Microbial strains of interest are genetically modified to constitutively produce a fluorescent protein. The emission spectra of the proteins must be sufficiently different in order to distinguish the strains on a Petri dish. To study selective sweeps, a mixture of the strains is prepared in liquid medium, usually with the fitter strain in the minority. A drop of this mixture is then deposited on a small region of a Petri dish with solid growth medium. Different shapes of this drop lead to different initial conditions. Circular drops are naturally created by the surface tension forces when small drops of fluid are placed on a Petri dish. Linear drops can be created by gently touching the surface of the medium with a razor blade after dipping it in the mixture of the strains. These drops dry quickly, and microbial colonies start to grow, expanding by about a centimeter a week. The spatial distribution of genotypes can be observed during this expansion by fluorescent microscopy, as shown in figure~\ref{FNSComparison}.

\begin{figure}
\includegraphics[width=8cm]{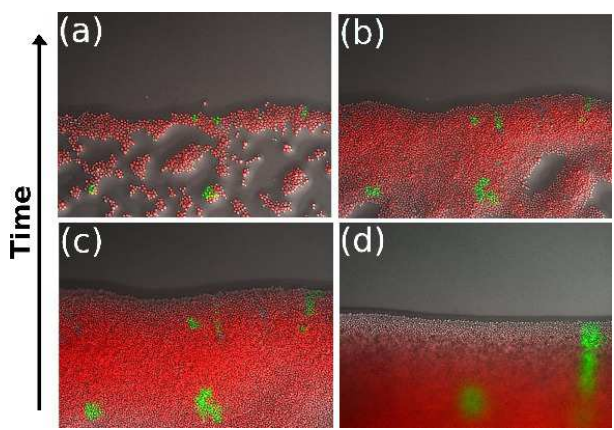}
\caption{(Colour online) Colony edge at single cell resolution~(mature yeast cells are $5\;\mu$m in diameter). (a), (b), (c), and (d) are successive images~(at two hour intervals) of the same region near the edge of a growing \textit{S. cerevisiae} colony inoculated with a razor blade. Note the formation of a green~(light gray) sector on the lower right. The two strains have approximately the same fitness in this experiment.}
\label{fig:high_resolution}
\end{figure}

Sectors in figure~\ref{FNSComparison} are at the center of this study. In our experiments, cells are nonmotile and grow primarily at the expanding frontier; behind the front, the growth is limited by the lack of nutrients~\cite{pipe:colonies,nadell:structure_plos}. As a result, the genetic composition in the interior of the colony does not change with time. The spatio-genetic pattern shown in figure~\ref{FNSComparison} is then a frozen record of the temporal changes in the spatial distribution of genotypes at the expanding frontier. In figure~\ref{FNSComparison}a, we show growth of two strains with the same fitness that differ only in the colour of a fluorescent marker. Before deposition on the Petri dish, these strains were combined in a 1:50 ratio and thoroughly mixed. However, the two different colours (genotypes) demix and form sectors. This demixing is caused by number fluctuations (genetic drift) at the expanding edge~\cite{HallatschekNelson:ExperimentalSegregation,Korolev:Review}. One can see this stochastic process at the resolution of a single cell~($5\;\mu$m in diameter) in figure~\ref{fig:high_resolution}. In contrast to neutral demixing, figure~\ref{FNSComparison}b shows sector formation in a colony founded by two strains with different fitnesses; the fitter strain is in the minority initially. Over time, the fitter strain displaces the other strain, increasing its share of the expanding frontier. This expansion is also subject to number fluctuations, which are responsible for the sector boundary wiggles, but the average shape of the sector is determined primarily by the deterministic force of natural selection. For the purpose of measuring relative fitness, the effects of genetic drift can be averaged out, given a sufficient number of repeated experiments. Our deterministic model strives to describe this average sector shape and is not capable of describing the randomness in boundary motion or the dynamics when the number of the fitter organisms is so small that the number fluctuations can lead to their extinction, an effect discussed in Ref.~\cite{Hallatschek:LifeFront}.

This paper is organized as follows. We formulate a competition model assuming a well-mixed environment, such as a mechanically shaken test tube, in Sec.~\ref{SCWellMixed}. Under these conditions, a particular microbe visits virtually every region of the carrier fluid in a cell division time, and the system is effectively ``zero-dimensional.'' This model is then extended to account for lateral migrations during a range expansion in Sec.~\ref{SCSpatial}. In Sec.~\ref{SCAnalysis}, we analytically derive the spatial patterns created by two-species competition~(including results for colliding circular colonies) using a very general argument, which does not rely on the microscopic details of microbial growth and migration. The theoretical predictions are then compared to experiments in Sec.~\ref{SComparison}. Concluding remarks are contained in Sec.~\ref{SConclusions}. The supplementary information (section~S1) contains the experimental and numerical methods as well as additional data supporting our conclusions.

\section{Modeling competition in a well-mixed environment}
\label{SCWellMixed}

A competition experiment is a standard way to measure relative fitness of two microbial strains in a well-mixed environment. During a competition experiment, the stains are introduced into a fresh medium, and their relative abundance is measured over time. Initially, the number of cells grows exponentially, but the growth eventually slows down as the system approaches the stationary phase due to crowded conditions. This behavior is captured by a simple Lotka-Volterra-type model~\cite{Murray:MathematicalBiology}:

\begin{equation}
\label{ECompetitionGeneral}
\left\{
\eqalign{
\frac{d}{dt}c_{1}(t)&=g_{1}c_{1}(t)-d_{11}c_{1}^{2}(t)-d_{12}c_{1}(t)c_{2}(t),\\
\frac{d}{dt}c_{2}(t)&=g_{2}c_{2}(t)-d_{21}c_{2}(t)c_{1}(t)-d_{22}c_{2}^{2}(t),\\}
\right.
\end{equation}

\noindent which is the most general model with quadratic nonlinearities. Here,~$c_{1}(t)$ and~$c_{2}(t)$ are the concentrations~(number of cells per unit volume) of strain~$1$ and strain~$2$ respectively. The constants~$g_{1}$ and~$g_{2}$ are their exponential growth rates; and the constant matrix~$d_{ij}>0$ describes nonlinear interactions. For a mono-culture consisting of a single yeast strain, Refs.~\cite{pearl:logistic,carlson:logistic} showed that population growth can be described by the logistic equation~(i.e. with quadratic nonlinearities) very accurately.

By rescaling~$c^{\rm{new}}_{1}(t)=d_{11}c_{1}(t)/g_{1}$ and~$c^{\rm{new}}_{2}(t)=d_{22}c_{2}(t)/g_{2}$, we can recast equation~(\ref{ECompetitionGeneral}) in a slightly more convenient form

\begin{equation}
\label{ECompetitionWellMixed}
\left\{
\eqalign{
\frac{d}{dt}c_{1}(t)&=g_{1}c_{1}(t)[1-c_{1}(t)-c_{2}(t)]+\epsilon_{1}c_{1}(t)c_{2}(t),\\
\frac{d}{dt}c_{2}(t)&=g_{2}c_{2}(t)[1-c_{1}(t)-c_{2}(t)]+\epsilon_{2}c_{2}(t)c_{1}(t),\\}
\right.
\end{equation}

\noindent where~$\epsilon_{1}=g_{1}-d_{12}g_{2}/d_{22}$ and~$\epsilon_{2}=g_{2}-d_{21}g_{1}/d_{11}$, and  we use the same symbols~$c_{1}(t)$ and~$c_{2}(t)$ for the rescaled concentrations. When~$c_{1}(t)+c_{2}(t)\approx1$, we shall say that the competition takes place ``under crowded conditions''. Our notation emphasizes that the initial stage of exponential growth is usually much shorter than the second phase of competition under crowded conditions, i.e. the growth rates~$g_{1}>0$ and~$g_{2}>0$, are typically much larger than the small quantities~$|\epsilon_{1}|$ and~$|\epsilon_{2}|$. When~$\epsilon_{i}>0$, strain~$i$ grows faster in the presence of the other strain, e.g. by feeding off an excess production of a useful amino acid. When~$\epsilon_{i}<0$, strain~$i$ grows slower in the presence of the other strain, e.g., because of a secreted poison.   

\begin{figure}
\psfrag{x}{\hspace{-1.0cm}$\bm{\epsilon_{1}>0,\;\epsilon_{2}>0}$}
\psfrag{y}{\hspace{-1.0cm}$\bm{\epsilon_{1}>0,\;\epsilon_{2}<0}$}
\psfrag{u}{\hspace{-1.0cm}$\bm{\epsilon_{1}<0,\;\epsilon_{2}>0}$}
\psfrag{v}{\hspace{-1.0cm}$\bm{\epsilon_{1}<0,\;\epsilon_{2}<0}$}
\includegraphics[width=8cm]{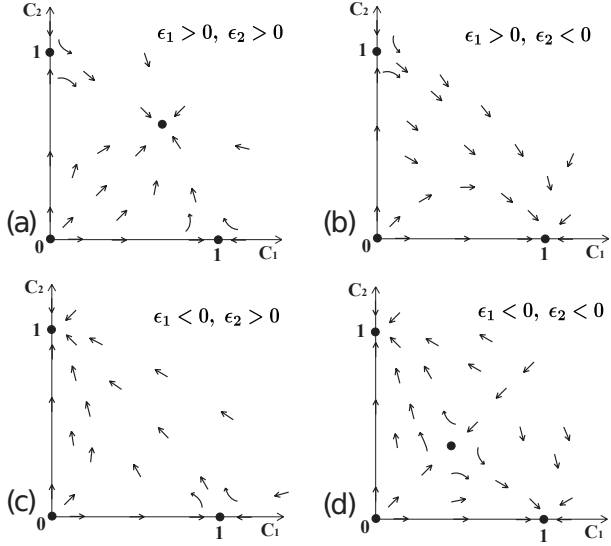}
\caption{Schematic representation of four different behaviors of equation~(\ref{ECompetitionWellMixed}) in phase space. (a)~$\epsilon_{1}>0$ and~$\epsilon_{2}>0$. (b)~$\epsilon_{1}>0$ and~$\epsilon_{2}<0$. (c)~$\epsilon_{1}<0$ and~$\epsilon_{2}>0$. (d)~$\epsilon_{1}<0$ and~$\epsilon_{2}<0$. The arrows represent the direction of trajectories in the phase space, and the dots represent the fixed points.}
\label{FWellMixedSchematicPP}
\end{figure}

Note that equations~(\ref{ECompetitionWellMixed}) always have at least three fixed points:~$(0,0)$,~$(1,0)$, and~$(0,1)$. In addition, for some values of the parameters, there is another fixed point~$(c_{1}^{*},c_{2}^{*})$ in the physically relevant domain of nonnegative~$c_{1}$ and~$c_{2}$. The fixed point at the origin is always unstable. 

Five different behaviors are possible depending on the values of~$\epsilon_{1}$ and~$\epsilon_{2}$; four of these are illustrated in figure~\ref{FWellMixedSchematicPP}. If~$\epsilon_{1}$ and~$\epsilon_{2}$ are positive, the interaction of the species is mutualistic~(i.e. the presence of the first strain helps the second strain grow and vice versa) and leads to a single stable fixed point~$(c_{1}^{*},c_{2}^{*})$ with a nonzero concentration of both strains. Since the focus of this paper is on competition, we do not pursue this cooperative possibility further here. If~$\epsilon_{1}$ and~$\epsilon_{2}$ are negative, there are two stable fixed points: one with~$c_{1}=1$ and~$c_{2}=0$, the other with~$c_{1}=0$ and~$c_{2}=1$. The system reaches one of these fixed points depending on which strain is more prevalent initially. An incoming separatrix divides the phase space in two domains of attraction and feeds into an unstable fixed point with~$c_{1}^{*}>0$ and~$c_{2}^{*}>0$. If~$\epsilon_{1}>0$ and~$\epsilon_{2}<0$, there is only one stable fixed point~$c_{1}=1$ and~$c_{2}=0$. Similarly,~$c_{1}=0$ and~$c_{2}=1$ is the only stable fixed point when~$\epsilon_{1}<0$ and~$\epsilon_{2}>0$. Finally, there is a degenerate case~$\epsilon_{1}=\epsilon_{2}=0$ when the dynamics is determined only by the logistical growth. In this case, depending on the initial conditions, the system lands somewhere along the line of neutral fixed points defined by~$c_{1}+c_{2}=1$.

\begin{figure}
\includegraphics[width=6cm]{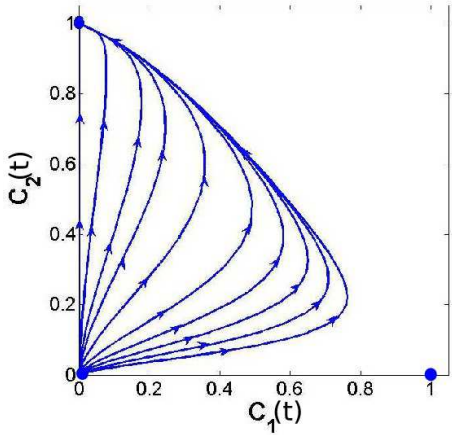}
\caption{(Colour online) Solutions of equation~(\protect{\ref{ECompetitionWellMixed}}) plotted in phase space~$(c_{1},c_{2})$ for negative~$\epsilon_{1}$ and positive~$\epsilon_{2}$. The arrows indicate the direction of time. For the initial conditions, we chose positive~$c_{1}(0)$ and~$c_{2}(0)$ close to zero with each trajectory having a different value of~$c_{1}(0)/c_{2}(0)$. In this plot,~$g_{1}=1.5$,~$g_{2}=1$,~$\epsilon_{1}=-0.1$,~$\epsilon_{2}=0.6$. Note that, initially, all trajectories bend toward the $c_{1}$-axis~(increasing~$c_{1}/c_{2}$), but the first strain is nevertheless eliminated at long times, as the system approaches the stable fixed point~$(0,1)$.}
\label{FWellMixedPP}
\end{figure}

Interestingly, when~$\epsilon_{1}\epsilon_{2}<0$, the ultimate result of the competition is independent of the exponential growth rates~$g_{1}$ and~$g_{2}$. The dynamical path, however, does depend on these parameters. For example, for~$\epsilon_{1}<0<\epsilon_{2}$ and~$g_{1}>g_{2}$~(the situation of particular interest to us here), $c_{1}(t)$~initially increases, but eventually falls off as the total population density rises settling at~$(c_{1},c_{2})=(0,1)$, see figure~\ref{FWellMixedPP}. This behavior has important implications for the competition at the frontier of a colony expanding in space, where the concentration of cells is always low. We show in Sec.~\ref{SCSpatial} that the outcome of a spatial competition experiment is determined primarily by the exponential growth rates~$g_{1}$ and~$g_{2}$ rather than~$\epsilon_{1}$ and~$\epsilon_{2}$, in contrast to the aforementioned behavior in a well-mixed population.

\section{Competition and migration}
\label{SCSpatial}

In this section, the competition model formulated in Sec.~\ref{SCWellMixed} is generalized to spatially extended environments. To this end, we need a model of cell migration. In general, cell migration is a complicated process, which could involve chemotaxis\footnote{Chemotaxis is an ability of cells to direct their motion in response to a chemical signal, e.g. food or toxins.}, swarming, and random wandering. Although all of these can be important biologically, some can be neglected in appropriately designed spatial competition experiments. For example, cells of Baker's yeast, \textit{Saccharomyces cerevisiae}, often used in microbiological experiments, are nonmotile, and many bacterial cells, e.g. \textit{Escherichia coli}, can be prevented from swimming by eliminating functioning flagella or using a high concentration of the gelation agent in the growth medium. 

For nonmotile microbes, the only mechanism of cell migration is cells' pushing on each other as they increase in size before cell division. Even a cell division of an isolated cell leads to migration because at least one of the offspring is generally displaced in a random direction relative to the position of its parent. At the colony front, where cells are relatively free to move, cell migration can be approximated by a random-walk-type process caused by growing cells pushing each other in random directions. The diffusion constant of such random walks must depend on the local concentration of cells because sector boundaries in the interior of a colony like the ones shown in figure~\ref{FNSComparison} do not change with time~\cite{HallatschekNelson:ExperimentalSegregation}. For a concentration-independent diffusion constant, the boundaries would slowly disappear as cells of different colours gradually mix~\cite{Hallatschek:LifeFront}; therefore, migration must be arrested at high cell concentrations. A possible mechanism of this arrest is a significant reduction in growth rate due to nutrient depletion. At the edge of the colony, however, the cell density is low, and cells move readily due to the jostling caused by cell growth, as is evident from the wandering of the sector boundaries.

Since the exact dependence of the spatial diffusion constant on the local cell concentration is unknown, we have explored a family of functional forms, namely

\begin{equation}
\label{SCDiffusionConstant}
\eqalign{
&D_{1}(c_{1},c_{2})=D_{01}(1-c_{1}-c_{2})^{\alpha_{1}} \;\; \mbox{for}\; c_{1}+c_{2}<1,\\
&D_{1}(c_{1},c_{2})=0 \;\; \mbox{for}\; c_{1}+c_{2}>1,}
\end{equation}

\noindent where~$D_{01}$ is the diffusion constant of the first strain in the limit of small local cell concentration, and~$\alpha_{1}$ is an adjustable parameter that allows us to explore the sensitivity to rapid variations in the rate of migration near the frontier. Here,~$c_{1}(t,\bm{x})$ and~$c_{2}(t,\bm{x})$ are cell densities per unit area, rescaled as in Sec.~\ref{SCWellMixed}. An analogous dependence,~$D_{2}(c_{1},c_{2})=D_{02}(1-c_{1}-c_{2})^{\alpha_{2}}\theta(1-c_{1}-c_{2})$, is assumed for the second strain; here~$\theta(x)$ is the Heaviside step function,~$\theta(x)=1,\; x\ge0$, and~$\theta(x)=0$ otherwise. The choice of a monotonic dependence of the diffusion constants on cell density is motivated by the monotonically decreasing supply of nutrients from the outside to the inside of the colony. To check whether a non-monotonic dependence would affect our results, we also explored~$D_{1}(c_{1},c_{2})=D_{2}(c_{1},c_{2})=D_{0}(c_{1}+c_{2})(1-c_{1}-c_{2})\theta(1-c_{1}-c_{2})$. Although the speed of population waves is different in this model, the shape of spatial patterns and their connection to relative fitness remain the same; see supplementary information (section~S4). 

Our model of spatial competition then takes the following form

\begin{equation}
\label{ECompetitionSpatialFull}
\left\{
\eqalign{
\frac{\partial c_{1}(t,\bm{x})}{\partial t}&=\bm{\nabla}\cdot\left[D_{1}(c_{1},c_{2})\bm{\nabla}c_{1}(t,\bm{x})\right]+\\ &g_{1}c_{1}(1-c_{1}-c_{2})+\epsilon_{1}c_{1}c_{2},\\
\frac{\partial c_{2}(t,\bm{x})}{\partial t}&=\bm{\nabla}\cdot\left[D_{2}(c_{1},c_{2})\bm{\nabla}c_{2}(t,\bm{x})\right]+\\ &g_{2}c_{2}(1-c_{1}-c_{2})+\epsilon_{2}c_{2}c_{1}.\\}
\right.
\end{equation}

\begin{figure}
\begin{tabular}{ll}
(a)\includegraphics[width=4cm]{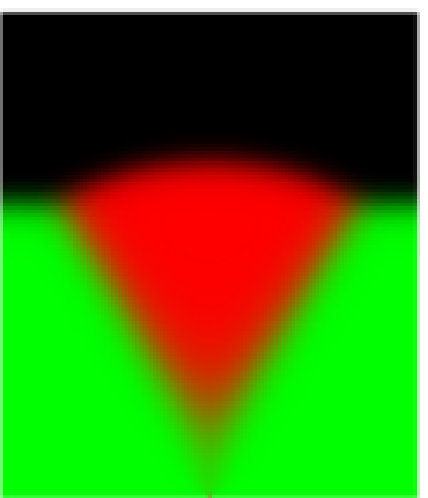} & (b)\includegraphics[width=4cm]{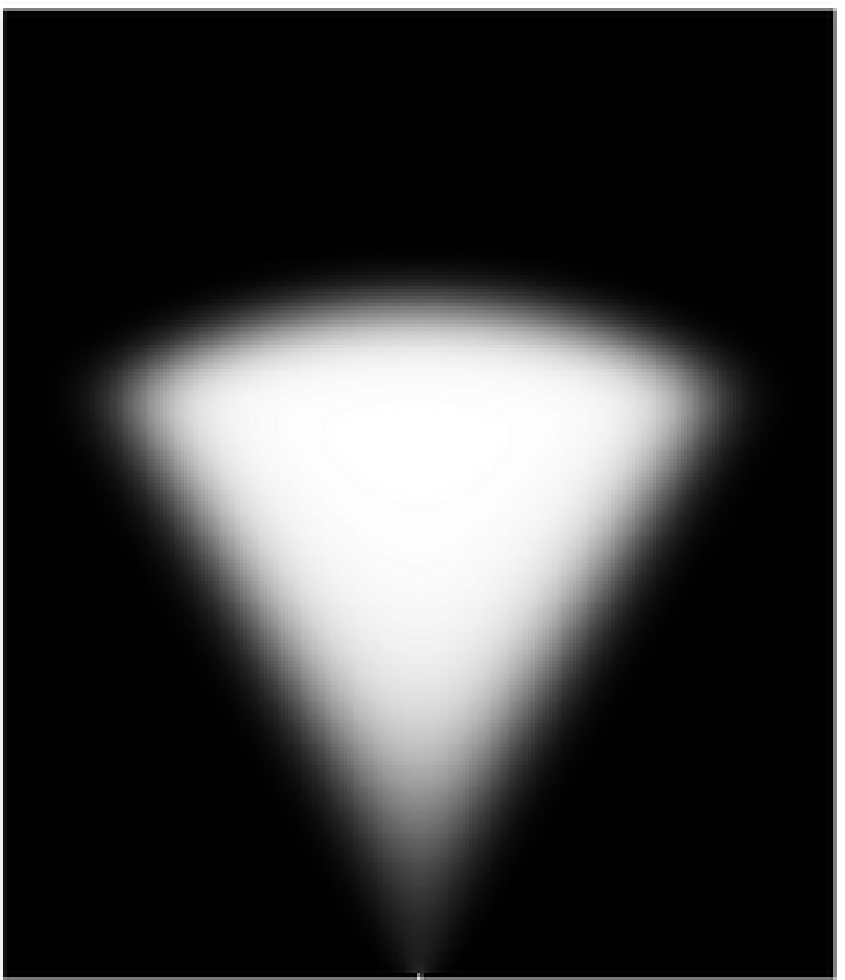}
\end{tabular}
\caption{(Colour online) Both (a) and (b) show the solution of equation~(\protect{\ref{ECompetitionSpatialFull}}) plotted for~$g_{1}=11.8$,~$g_{2}=10$,~$\epsilon_{1}=\epsilon_{2}=0$,~$\alpha_{1}=\alpha_{2}=1$, $D_{01}=1.18\cdot10^{-3}$ and~$D_{02}=10^{-3}$; note that we chose~$g_{1}/g_{2}=D_{01}/D_{02}=1.18$ to facilitate comparison with figure~\protect{\ref{FNSComparison}}b (see discussion in text). In these units, the habitat is a~$1\times2$ rectangle and is initially empty; only the bottom sixty percent of the habitat is shown because the top part remains empty throughout the expansion. The origin of the expansion is a line at the bottom edge of the images, where we impose the boundary conditions that~$c_{2}(t,x,0)=1$ and~$c_{1}(t,x,0)=0$ except in the width~$2^{-8}$ region near the center of this boundary, where~$c_{2}(t,x,0)=0$ and~$c_{1}(t,x,0)=1$. No-flux boundary conditions are imposed along all other edges. Equation~(\protect{\ref{ECompetitionSpatialFull}}) is solved on a square grid of~$256\times512$ points. (a) The concentration of the first strain is shown in red~(dark gray) and of the second strain in green~(light gray). The maximal colour intensity corresponds to the concentration of~$1$, and the lowest to the concentration of~$0$. This colour scheme is chosen to facilitate the comparison with the experimental data shown in figure~\protect{\ref{FNSComparison}}. (b) The same solution as in (a), but only the concentration of the first strain is shown to highlight  its establishment as a sector early in the expansion. Brighter regions correspond to higher concentration of the first~(red) strain.}
\label{FSCIllustrationSolution}
\end{figure}

\begin{figure}
\includegraphics[width=6cm]{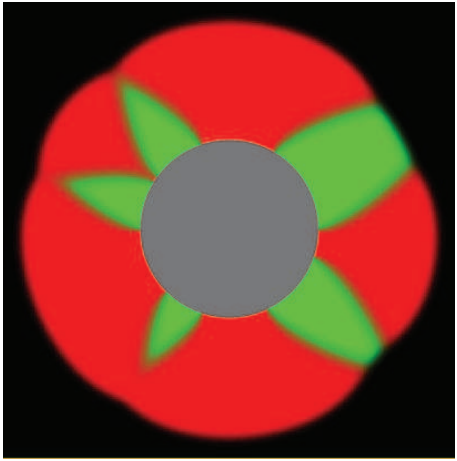}
\caption{(Colour online) The solution of equation~(\ref{ECompetitionSpatialFull}) plotted for~$g_{1}=12.5$,~$g_{2}=10$,~$\epsilon_{1}=\epsilon_{2}=0$,~$\alpha_{1}=\alpha_{2}=1$, and~$D_{01}=D_{02}=0.01$. The habitat is a~$10\times10$ square inoculated with a circular drop of radius~$2$ shown in gray. We assign~$c_{1}$ and~$c_{2}$ in the initial circular boundary to be either~$0$ or~$1$~(in blocks) to mimic the sectoring pattern produced by a short period of genetic drift with a relatively small selective advantage. No-flux boundary conditions are imposed along all edges. Equation~(\protect{\ref{ECompetitionSpatialFull}}) is solved on a square grid of~$2560\times2560$ points. The concentration of the first strain is shown in red and of the second strain in green. The maximal intensity~(of red or green) corresponds to the concentration of~$1$, and the lowest to the concentration of~$0$.}
\label{FSCICircularSolution}
\end{figure}

\noindent The initial and boundary conditions are chosen to mimic selective sweeps in microbial colonies. For example, in figure~\ref{FSCIllustrationSolution}, only a very small region is initially occupied by the advantageous genotype, which corresponds to the time when a sector begins to appear in figure~\ref{FNSComparison}b. Our model cannot describe the earlier part of the range expansion when number fluctuations are important, because equations~(\ref{ECompetitionSpatialFull}) are deterministic and treat the cell densities as continuous functions. During this early stage, the advantageous genotype becomes extinct stochastically everywhere but a few spatial locations, where it gives rise to small sectors. These sectors can then be used as the initial condition in our deterministic model. We solve equations~(\ref{ECompetitionSpatialFull}) numerically; see the supplementary information (section~S1) and figures~\ref{FSCIllustrationSolution} and~\ref{FSCICircularSolution}. Note that our model of range expansions has two spatial dimensions, while, in experiments, colonies also gradually thicken in the direction perpendicular to the plate~\cite{nguyen:instability}. We neglect this gradual thickening here.

\begin{figure}
\includegraphics[width=8cm]{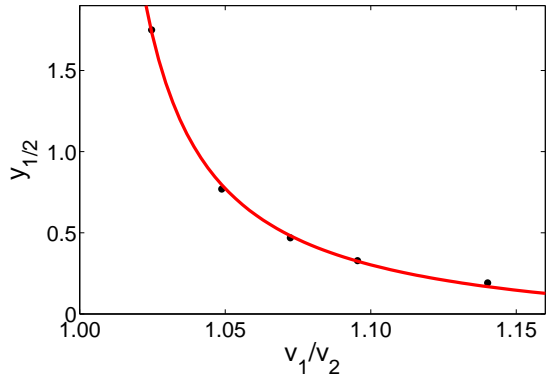}
\caption{(Colour online) The length of the initial stage of interacting sector boundaries from equation~(\ref{ECompetitionSpatialFull}) as a function of the ratio of expansion velocities~$v_{1}/v_{2}$. The quantity~$y_{1/2}$ is the distance from the origin of the sector to the closest point where~$c_{1}=1/2$. Here, we vary~$g_{1}$ while keeping~$D_{0}=10^{-3}$ and~$g_{2}=10$ fixed. For small fitness differences, we expect~$y_{1/2}\sim(g_{1}-g_{2})^{-1}$ from equation~(\ref{ETransitorySolution}). The data from numerical solutions of equation~(\ref{ECompetitionSpatialFull}) is shown as dots, and the solid line is a fit to~$A/(g_{1}-g_{2})+B$, where~$A$ and~$B$ are fitting parameters.}
\label{FTransitoryStage}
\end{figure}

From figures~\ref{FNSComparison}b and~\ref{FSCIllustrationSolution}, one can see that there are at least two stages in the sector formation. During the late stage, the two sector boundaries are far apart, and the interior of the sector is dominated by the advantageous strain. At this stage, the sector boundaries maintain a constant angle with the direction of the expansion, which we explain in Sec.~\ref{SCAnalysis} using a very general geometric argument. During the early stage, the size of the sector is comparable to the width of sector boundaries, and the two boundaries interact. By neglecting the nonlinear terms in equations~(\ref{ECompetitionSpatialFull}) near the frontier, where~$c_{1}$ and~$c_{2}$ are small, we can qualitatively understand how the duration of the first stage depends on the relative fitness when the fitness difference is small~(see figure~\ref{FTransitoryStage}). In this case, we can neglect the bulging of the sector and assume that the population wave front is approximately flat. For simplicity, we also assume~$D_{01}=D_{02}=D_{0}$. The~$x$-axis is taken to be along the front and the~$y$-axis to be perpendicular to the front. For the Fisher equation~\cite{Fisher:FisherWave}, a steady state is reached in a frame co-moving with the population wave~($x'=x$ and~$y'=y-2\sqrt{g_{2}D_{0}}t$) when only the second strain is present. In this reference frame,~$g_{2}c_{2}\approx-D_{0}\frac{\partial^{2}c_{2}}{\partial y'^{2}}-v_{2}\frac{\partial c_{2}}{\partial y'}$ (note that the nonlinear terms are neglected). Since the fitness difference and the concentration of the first strain are small, the dynamics of the second strain along the $y$-axis is approximately unchanged. Moreover, since the spatial distribution of the first strain along the~$y$-axis is the same as that of the second strain, a similar equality holds for the first strain~$g_{2}c_{1}\approx-D_{0}\frac{\partial^{2}c_{1}}{\partial y'^{2}}-v_{2}\frac{\partial c_{1}}{\partial y'}$. Upon using these two observations and the linearized version of equations~(\ref{ECompetitionSpatialFull}), we find that the dynamics of the first strain near the frontier is approximately given by a linear diffusion equation with a source, 

\begin{equation}
\label{ECompetitionSpatialSimplifiedComoving}
\frac{\partial c_{1}(t,x',y')}{\partial t}=D_{0}\frac{\partial^{2}c_{1}(t,x',y')}{\partial x'^{2}}+(g_{1}-g_{2})c_{1}(t,x',y'),
\end{equation}

\noindent Note that this equation is invariant with respect to translations along~$y'$, so we can treat the frontier as a quasi-one-dimensional population. For~$c_{1}(0,x',y')=\delta(x')$, corresponding to a point-like inoculant at the frontier, the solution of equation~(\ref{ECompetitionSpatialSimplifiedComoving}) is

\begin{equation}
\label{ETransitorySolution}
c_{1}(t,x')=\frac{1}{\sqrt{4\pi D_{0}t}}e^{(g_{1}-g_{2})t}e^{-\frac{x'^{2}}{4D_{0}t}}.
\end{equation}

\noindent Therefore, the characteristic time necessary for the first strain to dominate the sector scales as~$(g_{1}-g_{2})^{-1}$. This divergent time scale is indeed observed in the numerical solutions; see figure~\ref{FTransitoryStage}.

The dependence of spatial patterns during two-species competition on various parameters in equation~(\protect{\ref{ECompetitionSpatialFull}}) was investigated in the context of linear expansions. We varied the exponents~$\alpha_{1}$ and~$\alpha_{2}$ by factors of~$2$, $\alpha_{1}=\alpha_{2}\in\{1/4,1/2,1,2,4\}$; the diffusion constants~$D_{01}$ and~$D_{02}$ by factors of~$10$, $D_{01}=D_{02}\in\{10^{-2},10^{-3},10^{-4}\}$; and the growth rates~$g_{1}$ and~$g_{2}$ by factors of~$10$, $g_{1}\in\{1,10,10^{2}\}$ with~$g_{1}/g_{2}\in\{1,1.01,1.05,1.1,1.3,1.5,2\}$. The competition parameters~$\epsilon_{1}$ and~$\epsilon_{2}$ were independently varied relative to~${\rm min}\{g_{1},g_{2}\}$, $\epsilon_{1},\epsilon_{2}\in[-0.5,0.5]{\rm min}\{g_{1},g_{2}\}$. This numeric exploration helped us identify important parameter combinations that control the shape of spatial patterns. We now turn to the discussion of these results.

From numerical analysis, we made an important observation that the expansion velocity of a strain growing in the absence of the other strain depends \textit{only} on the exponential growth rate and diffusion constant right at the frontier and is given by

\begin{equation}
\eqalign{
&v_{1}=2\sqrt{D_{01}g_{1}},\\
&v_{2}=2\sqrt{D_{02}g_{2}},\\}
\label{EFisherVelocityFull}
\end{equation}

\noindent independent of~$\epsilon_{1},\;\epsilon_{2},\;\alpha_{1},\;\alpha_{2}$ (see supplementary information section~S4) and in agreement with the classic Fisher-Kolmogorov wave theory~\cite{Fisher:FisherWave,Kolmogorov:FKPPEquation}, which provides an exact solution for a simpler model with concentration-independent diffusivity. This agreement is not surprising because the speed of a Fisher population wave is determined only by the dynamics at the foot of the wave front~\cite{Fisher:FisherWave,Murray:MathematicalBiology}, and equation~(\ref{SCDiffusionConstant}) ensures that the diffusivity approaches a constant for small~$c_{1},\;c_{2}$. The intuition behind equation~(\ref{EFisherVelocityFull}) is that the wave speed depends both on the growth rate~($g_{1}$) and on the rate of undirected migration~($D_{01}$) that brings cells to unoccupied territories. The detailed shape of the wave front, however, does depend on parameters such as~$\alpha_{1}$ and~$g_{1}/D_{01}$~(or~$\alpha_{2}$ and~$g_{2}/D_{02}$). The parameters~$\epsilon_{1}$ and~$\epsilon_{2}$ are irrelevant when only one strain is present because~$c_{1}c_{2}=0$ in this case.

Even when both strains are present, the knowledge of expansion velocities~$v_{1}$ and~$v_{2}$ is sufficient to describe the major (large scale) features of the resulting spatio-genetic pattern. To see this, note that the behavior of population fronts far from the sector boundaries is the same as when only one strain is present because the concentration of the other strain vanishes away from the boundaries. We also found that the initial position of the sector boundaries is determined only by~$v_{1}$ and~$v_{2}$ (and independent of~$\epsilon_{1}$ and~$\epsilon_{2}$) because, at the tip of the advancing front, the product~$c_{1}c_{2}$ is exponentially small compared to~$c_{1}$ and~$c_{2}$. On smaller length scales of the order~$\sqrt{D_{01}/g_{1}}$~(or~$\sqrt{D_{02}/g_{2}}$), all parameters play a role. In particular, all parameters affect the shape of concentration profiles and the position of sector boundaries during the early stage of sector formation. 

\begin{figure}
\includegraphics[width=8cm]{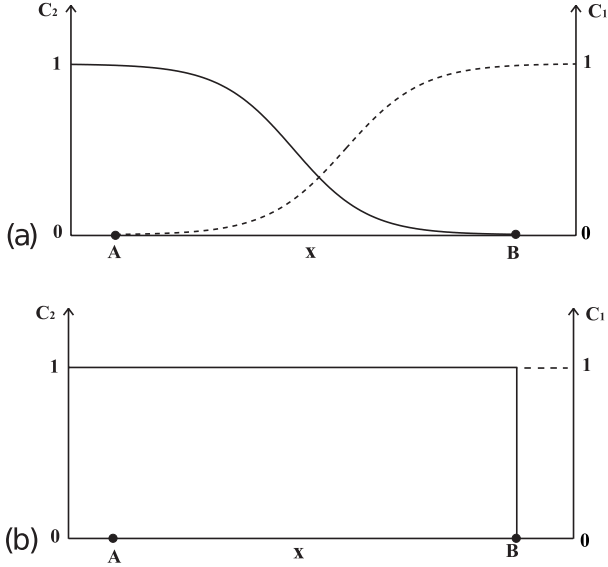}
\caption{Schematic illustration of transverse boundary motion behind a front advancing in the $y$-direction for~$g_{1}>g_{2}$ and~$\epsilon_{1}<0<\epsilon_{2}$. The plots show concentration profiles along a linear cut along the $x$-direction, parallel to the front and inoculant. (a) The concentration profiles a short distance behind the population frontier. There is an overlap region, where both~$c_{1}(x)$~(dashed line) and~$c_{2}(x)$~(solid line) are nonzero. This region has a finite width~(from A to B) because the discreetness of the number of cells is inconsistent with infinitesimally small values of the concentrations. (See Ref.~\protect{\cite{Hallatschek:FisherWave}} for a more detailed discussion of this issue in a related model.) (b) The concentration profile at the same spatial location as in (a), but after a very long time. The interval between A and B is now occupied exclusively by the cells of the second strain, which wins out under crowded conditions.}
\label{FBehindFront}
\end{figure}

Note that, with nonzero~$\epsilon_{1}$ and~$\epsilon_{2}$, sector boundaries still move behind the wave front even though the diffusivity is zero. To understand this motion, consider strains with~$g_{1}>g_{2}$ and~$\epsilon_{1}<0<\epsilon_{2}$. The first strain expands faster, but, after the front has passed, the dynamics under crowded conditions, discussed in Sec.~\ref{SCWellMixed}, favors the second strain. As a result, any region with a nonzero concentration of the second strain is eventually colonized by it. This behind-the-front competition should lead to a finite displacement of the boundary because every sector boundary has a finite width due to the discreteness of the number of organisms; see figure~\ref{FBehindFront}.  Since, for many microbial strains, the boundary width is small, and~$g_{1},g_{2}\gg|\epsilon_{1}|,|\epsilon_{2}|$, we do not expect to observe this type of sector boundary displacement~(very different from a Fisher genetic wave) experimentally.

What is the relation between~$v_{1}/v_{2}$ and relative fitness in liquid cultures? In well-mixed populations, selective advantage is often defined from the ratio of exponential growth rates,~$s_{\rm{wm}}=g_{1}/g_{2}-1$. At expanding frontiers, we define~$s=v_{1}/v_{2}-1$ by analogy. There may not be a direct correspondence between~$s$ and~$s_{\rm{wm}}$ because the former involves diffusion constants~$D_{01}$ and~$D_{02}$ in addition to the growth rates~$g_{1}$ and~$g_{2}$. For example, a mutation providing a means of motility could be beneficial in a Petri dish (due to faster spreading) and deleterious in liquid (due to its metabolic cost). However, two special cases~$D_{01}=D_{02}$ and~$D_{01}\sim g_{1}$ (with $D_{02}\sim g_{2}$) are of interest.

\begin{figure}
\begin{tabular}{ll}
\includegraphics[width=6cm]{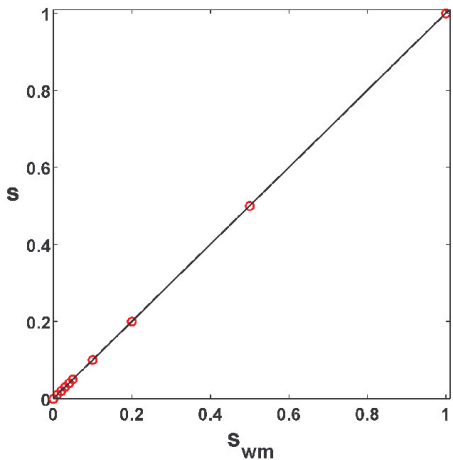} 
\end{tabular}
\caption{(Colour online) The comparison between selective advantage in liquid culture and on Petri dishes within the reaction-diffusion model. The red circles show the results of the numerical solution of equation~(\protect{\ref{ECompetitionSpatialFull}}) for single strain expansions. The black line shows the theoretically predicted linear dependence. We varied~$g_{1}$ to mimic different growth rates in the experiments and used~$g_{2}=10$,~$\epsilon_{1}=\epsilon_{2}=0$,~$\alpha_{1}=\alpha_{2}=1$,~$D_{01}=10^{-4}g_{1}$, and~$D_{02}=10^{-3}$. In these units, the habitat was a~$1\times10$ rectangle and was initially empty. Each expansion was started at the shorter edge of the habitat, where we imposed the Dirichlet boundary condition forcing strain density to be~$1$. No-flux (Neumann) boundary conditions were imposed along all other edges. The grid size used was~$128\times1280$ points.}
\label{fig:fitness_plate_liquid}
\end{figure}

Equal diffusion constants should be a good approximation for mutations that affect growth rate, but do not affect motility directly, which is possible when motility does not depend on the growth rate strongly. Growth independent motility was, e.g., observed in swimming \textit{Bacillus subtilis}  cells~\cite{wakita:expansion}. Under these assumptions,~$1+s_{\rm{wm}}=(1+s)^{2}$, and~$s_{\rm{wm}}=2s$ for~$s\ll1$; see equation~(\ref{EFisherVelocityFull}). The other possibility~$D_{0i}\sim g_{i}$ could be a good approximation when motility and growth are strongly linked. For example, colonies of~\textit{S. cerevisiae} studied here expand due to cell growth, and it is reasonable to assume that~$D_{0i}\sim g_{i}a^{2}$, where~$a$ is the average cell size. In this case,~$s_{\rm{wm}}=s$, as follows from equation~(\ref{EFisherVelocityFull}); see figure~\ref{fig:fitness_plate_liquid}.

Our simple definition of~$s$~($s=v_{1}/v_{2}-1$) has three advantages: expansion velocities can be easily measured, larger expansion velocity results in greater colonized territory and, therefore, greater access to nutrients, and the ratio~$v_{1}/v_{2}$ is closely related to the traditional definition of~$s_{\rm{wm}}$ obtained from the exponential phase of well-mixed cultures.

In summary, the main conclusion of the mechanistic modeling embodied by equation~(\ref{ECompetitionSpatialFull}) is that the essential features of spatio-genetic patterns formed during range expansion are insensitive to the details of the model and are determined by a single dimensionless parameter~$v_{1}/v_{2}$~(the ratio of expansion velocities). We further support this conclusion in the next section by deriving the shapes of the spatial pattern without relying on the microscopic dynamics of growth and migration. Additional tests of the robustness of our reaction-diffusion model to changes in the modeling assumptions~(such as varying the form of the concentration-dependent diffusion constants) are presented in section~S4 of the supplementary materials.

\section{Sector shapes and the equal-time argument}
\label{SCAnalysis}

In this section, we develop an analytic argument to understand selective sweeps and explore different ways of measuring relative fitness from macroscopic competition experiments. This argument relies on two assumptions: the strains expand with constant velocities and a patch occupied by one strain is impenetrable to the other strain. We primarily focus on two experimental geometries: linear and circular. 

\subsection{Linear inoculations}

\begin{figure}
\includegraphics[width=8cm]{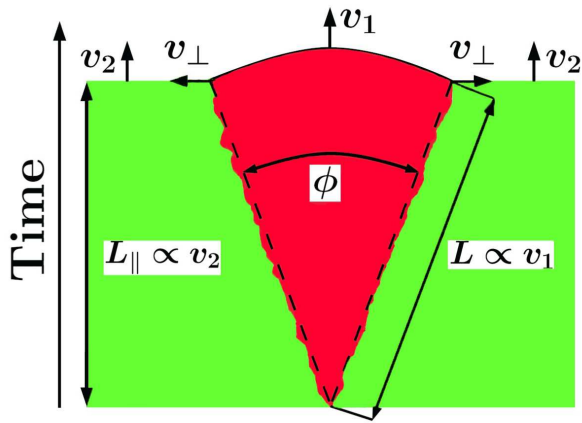}
\caption{(Colour online) Equal-time argument and sector shape in a linear geometry. The wiggles in the sector boundaries represent genetic drift, neglected in most of this paper.}
\label{FLinearGeometry}
\end{figure}

Upon establishment, i.e. when sector boundaries are sufficiently far apart, the sectors from a linear inoculation have a triangular shape. Following Ref.~\cite{Hallatschek:LifeFront}, we explain this shape by the simple equal-time argument illustrated in figure~\ref{FLinearGeometry}: It should take the same amount of time for the second~(less fit) strain to grow along the expansion direction as it takes the fitter first strain to grow a longer distance along the boundary. More generally, for any point along the front, we can define a length~$\rho$ that is the length of the shortest path connecting this point and the origin of the expansion and lying entirely in the territory occupied by the same strain. (This path is a straight line for linear inoculations, but curved for radial ones, as we show below.) Then, the ratio of~$\rho$ and the appropriate expansion velocity is the time necessary to form this particular spatial pattern. This time is the age of the colony since inoculation and must be the same for all points along the front. From this observation~(see figure~\ref{FLinearGeometry}), we conclude that the bulging shape of an advantageous sector is an arc of a circle of radius~$v_{1}t$ and angle~$\phi$ given by

\begin{equation}
\label{EPhiET}
\tan\left(\frac{\phi}{2}\right)=\sqrt{\frac{v_{1}^{2}}{v_{2}^{2}}-1}=\sqrt{s(2+s)},
\end{equation}

\noindent which is equivalent to the result obtained in Ref.~\cite{Hallatschek:LifeFront}.

Our arguments are somewhat analogous to the Huygens-Fresnel principle in optics~\cite{Born:Optics}. Each point along the front has a potential to create an outgoing circular population wave that spreads with velocity~$v_{1}$ or~$v_{2}$, and unoccupied territories are colonized by the strain that gets there first. Unlike in optics, this principle can only be used to construct wave fronts at infinitesimal time steps because, once a region is colonized by one strain, it becomes impenetrable to the other strain.

The equal-time argument breaks down on length scales smaller than~$D_{0}/v_{1}\propto \sqrt{D_{0}/g_{1}}$ because the Fisher velocity will in general depend on the curvature of the wave front. For example, advancing a circular front requires more time than advancing a linear front because more cell divisions are necessary to cover the larger area colonized by the curved front. The Fisher velocity approaches the limiting value, equation~(\ref{EFisherVelocityFull}), provided the local radius of curvature is much larger than~$D_{0}/v_{1}$~\cite{Murray:MathematicalBiology}. $D_{0}/v_{1}$ is also the characteristic width of a sector boundary, where the equal-time argument breaks down because the strains are intermixed.

\subsection{Circular expansions}

We now turn to radial range expansions, resulting from moderate-sized circular pioneer populations, e.g., created by placing a drop of the inoculant on the surface of a Petri dish. Due to  surface tension, small drops~(of order~$5\;\mu\rm{l}$ to~$3\;$mm in diameter) are naturally circular. A precisely defined initial shape is a significant advantage over linear geometries susceptible to front undulations. This advantage is further strengthened because a two-dimensional reaction-diffusion population wave started from an irregular island of cells becomes more and more circular as the expansion continues. Even more important, the interactions of yeast cells lead to an effective surface tension suppressing front undulations~\cite{nguyen:instability}.

\begin{figure}
\includegraphics[width=6cm]{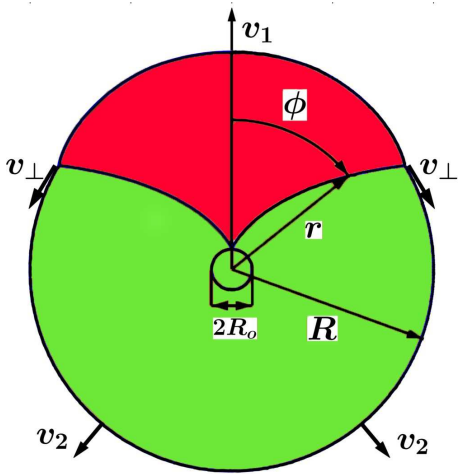}
\caption{(Colour online) Equal-time argument and sector shape for a circular inoculant of radius~$R_{0}$.}
\label{FCircularGeometry}
\end{figure}

The equal-time argument also yields the shape of sectors in the circular geometry; see the schematic plot~(figure~\ref{FCircularGeometry}) and numerical solution of equation~(\ref{ECompetitionSpatialFull}) (figure~\ref{FSCICircularSolution}). Equating the infinitesimal time increments along the radius of the wild-type colony and the curved sector boundary leads to a differential equation:
\begin{equation}
\label{ECircularSectorEquation}
\frac{dr}{v_{2}}=\frac{\sqrt{dr^{2}+(rd\phi)^{2}}}{v_{1}},
\end{equation}

\noindent formulated in polar coordinates~$(r,\phi)$ with the origin at the center of the expansion. The solution reads

\begin{equation}
\label{ECircularSectorSolution}
\phi=\pm\sqrt{\frac{v_{1}^{2}}{v_{2}^{2}}-1}\ln\left(\frac{r}{R_{0}}\right)=\pm\sqrt{s(2+s)}\ln\left(\frac{r}{R_{0}}\right),
\end{equation}

\noindent where the different signs corresponds to boundaries turning clockwise and counterclockwise, and $R_{0}$ is the initial radius of the population. Thus, we can identify sector boundaries in microbiology with the famous logarithmic spiral of Bernoulli, which also describes the Nautilus shell and insect flight patterns~\cite{cook:curves_life}.

\begin{figure}
\includegraphics[width=7cm]{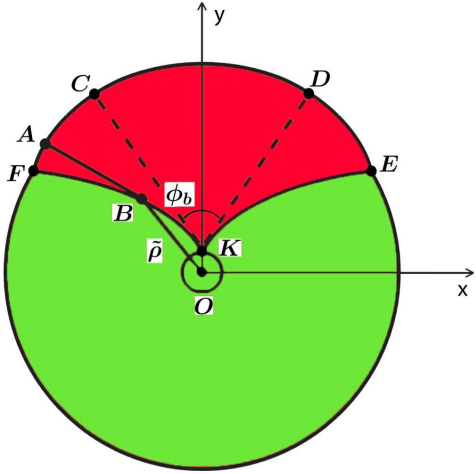}
\caption{(Colour online) Illustration of the equal-time argument for the shape of the bulge induced by a faster growing strain~(strain~$1$) in the circular geometry; see also figure~\ref{FCircularGeometry}. The central part of the bulge~(between~$C$ and~$D$) is an arc of a circle bounded by two tangents~$CK$ and~$DK$~(black dashed lines) to the sector boundaries at their origin. The rest of the bulge~(between~$F$ and~$C$, and between~$D$ and~$E$) is described parametrically by equation~(\protect{\ref{EBulgeAngleShape}}), where, for any point~$A$ on this part of the bulge, the parameter~$\tilde{\rho}$ is the distance between the center of the homeland and the intersection point~$B$ between the closest sector boundary and its tangent~$AB$ passing through~$A$. Equating the total expansion time of the fitter strain, first, along the sector boundary~$KB$ and, then, along the tangent~$AB$ to the expansion time of the other strain along the radius of the green segment immediately yields equation~(\protect{\ref{EBulgeAngleShape}}).\label{FBulgeShape}}
\end{figure}

The shape of the bulge at the frontier can also be calculated with the equal-time argument. The top of the bulge, well away from the sector boundaries, is a circular arc of radius~$v_{1}t$ centered on the origin of the sector; the angular length~$\phi_{b}$ of this region~(see figure~\protect{\ref{FBulgeShape}}) is given by

\begin{equation}
\label{EBulgeAngleCircle}
\tan\left(\frac{\phi_{b}}{2}\right)=\sqrt{\frac{v_{1}^{2}}{v_{2}^{2}}-1},
\end{equation}

\noindent just as in the linear geometry. Beyond~$\phi_{b}/2$, the bulge is closer to the sector origin than~$v_{1}t$ because these points along the bulge cannot be connected to the origin of the sector by a straight line without intersecting the territories occupied by the other strain. In this case, the shortest allowed path back to the founding population is a straight line passing through a specific point~(point~$B$ in figure~\ref{FBulgeShape}) followed by a curved path along the sector boundary to the origin of the sector. From figure~\ref{FBulgeShape}, one can easily see that the straight path is tangent to the sector boundary. Upon invoking the equal-time argument~(see figure~\ref{FBulgeShape}), we find that the shape of the bulge beyond~$\phi_{b}/2$ is described by

\begin{equation}
\left\{
\eqalign{
x&=\tilde{\rho}\sin\kappa+(R_{0}+v_{2}t-\tilde{\rho})(\sin\kappa+\sqrt{\frac{v_{1}^{2}}{v_{2}^{2}}-1}\cos\kappa),\\
y&=\tilde{\rho}\cos\kappa+(R_{0}+v_{2}t-\tilde{\rho})(\cos\kappa-\sqrt{\frac{v_{1}^{2}}{v_{2}^{2}}-1}\sin\kappa),\\}
\right.
\label{EBulgeAngleShape}
\end{equation}

\noindent where~$t=0$ is the time of the inoculation,~$\kappa$ is given by

\begin{equation}
\label{Ekappa}
\kappa=\sqrt{\frac{v_{1}^{2}}{v_{2}^{2}}-1}\ln\left(\frac{\tilde{\rho}}{R_{0}}\right),
\end{equation}

\noindent and~$\tilde{\rho}\in(R_{0},R_{0}+v_{2}t)$ is a parameter equal to the length of~$OB$ in figure~\ref{FBulgeShape}. 

\begin{figure}
\includegraphics[width=8cm]{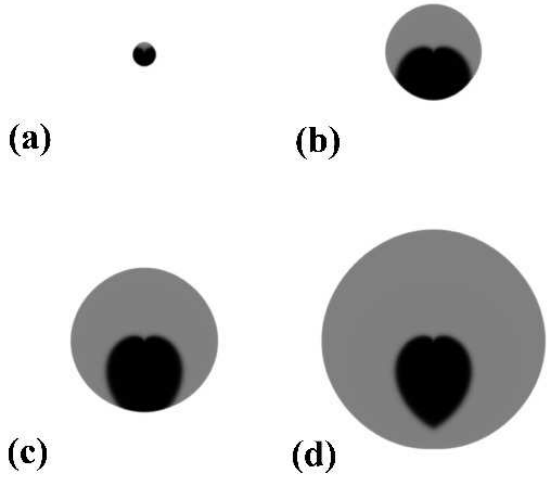}
\caption{The enclosure of a slower growing strain~(initially in the majority) by the faster growing strain during a competition experiment. The fitter strain is shown in gray, and the other strain is shown in black. Four consecutive snapshots of the numerical solution of equation~(\protect{\ref{ECompetitionSpatialFull}}) are shown in (a), (b), (c), and (d). The enclosure leading to the heart-shape occurs shortly before the snapshot shown in (d).}
\label{FHeart}
\end{figure}

Since the two sector boundaries turn in opposite directions, they must eventually meet, thus, enclosing the less fit strain within the population of a faster growing strain, as shown in figure~\ref{FHeart}. This enclosure occurs at~$\phi=\pm\pi$, and, from equation~(\ref{ECircularSectorSolution}), we calculate the distance~$R_{f}$ from the center of the inoculation to the point where the two boundaries finally meet,

\begin{equation}
\label{EBulgeAngleCircle2}
R_{f}=R_{0}\exp\left(\frac{\pi v_{2}}{\sqrt{v_{1}^{2}-v_{2}^{2}}}\right)\approx R_{0}\exp\left(\frac{\pi}{\sqrt{s(2+s)}}\right),
\end{equation}

\noindent where the last equality follows from our definition of selective advantage~$v_{1}=(1+s)v_{2}$. Note that the time~$t_{f}=R_{f}/v_{2}$ at which the second strain is enclosed by the first strain is exponentially large as~$s\to0$. Therefore, for small~$s$, a competitive exclusion requires a much longer time than~$(g_{1}-g_{2})^{-1}=1/s_{\rm{wm}}$ predicted by the well-mixed population models. 

Equation~(\ref{EBulgeAngleShape}) is valid only up to~$t=R_{f}/v_{2}$, and, after the enclosure, the shape of the colony is determined by the expansion of the first strain with velocity~$v_{1}$.

Another interesting consequence of the equal-time argument is that the point where the two strains meet at the colony edge moves tangentially to the front of the less fit strain with a constant velocity~$v_{\perp}=\sqrt{v_{1}^{2}-v_{2}^{2}}$; see figures~\ref{FLinearGeometry} and~\ref{FCircularGeometry}. Indeed, the distance~$l(t)$ between the two sector boundaries at a linear front increases with time as~(equation~\ref{EPhiET})

\begin{equation}
\eqalign{
l(t)&=l(t_{0})+2v_{2}(t-t_{0})\sin(\phi/2)\\
&=l(t_{0})+2\sqrt{v_{1}^{2}-v_{2}^{2}}(t-t_{0}),}
\label{eq:l_linear}
\end{equation}

\noindent and the angular size of a sector~$\varphi(t)$ in the circular geometry grows as~(equation~\ref{ECircularSectorSolution})

\begin{equation}
\varphi(t)=\varphi(t_{0})+\frac{2\sqrt{v_{1}^{2}-v_{2}^{2}}}{v_{2}}\ln\left[\frac{R(t)}{R(t_{0})}\right],
\label{eq:l_circular}
\end{equation}

\noindent where the factors of 2 are due to boundary motion at both edges of a sector. The lateral expansion velocity is then given by

\begin{equation}
v_{\perp}=\frac{1}{2}\frac{dl(t)}{dt}=\frac{1}{2}R(t)\frac{d\varphi(t)}{dt}=\sqrt{v_{1}^{2}-v_{2}^{2}}.
\label{eq:v_perp}
\end{equation}

\noindent Thus,~$v_{\perp}$ is a constant. One can also show that~$\vec{v}_{\perp}=\vec{v}_{1}-\vec{v}_{2}$ and these three vectors make a right triangle~($\vec{v}_{\perp}\perp\vec{v}_{2}$); here~$\vec{v}_{1}$ and $\vec{v}_{2}$ are front velocities at the sector boundary. Sectors of deleterious strain also obey equation~(\ref{eq:v_perp}), but with a negative lateral velocity,~$v_{\perp}=-\sqrt{|v_{1}^{2}-v_{2}^{2}|}$.

\subsection{Colony collisions}

\begin{figure}
\includegraphics[width=7cm]{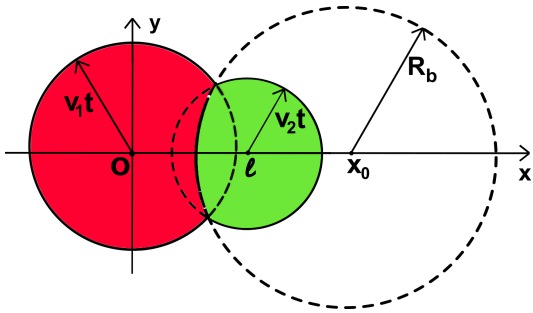}
\caption{(Colour online) Schematic picture of a colony collision.}
\label{fig:collision_theory}
\end{figure}

We conclude this section by considering competition between two strains not initially in contact. At the beginning of the experiment, two circular colonies inoculated with different strains grow independently each with their own velocity. Eventually, however, the colonies collide; see figure~\ref{fig:collision_theory}. For simplicity, we assume that the initial radius of the colonies is much smaller than the distance~$l$ between the colonies. Provided the nutrients remain abundant by time~$t$, the colony boundaries are circles with radii~$v_{1}t$ and~$v_{2}t$ except for the collision boundary. The shape of the collision boundary follows from the equal-time argument; see figure~\ref{fig:collision_theory}. Remarkably, this boundary is \textit{also} a circle, with radius~$R_{b}$ given by

\begin{equation}
R_{b}=l\frac{v_{1}v_{2}}{|v_{1}^{2}-v_{2}^{2}|}=l\frac{1+s}{s(2+s)}.
\label{eq:Rb}
\end{equation} 

\noindent Note that~$R_{b}$ diverges as the selective advantage~$s\to0$. The center of this circle is located on the line connecting the centers of the colonies, distance~$x_{0}$ away from the center of the colony established by the first~(faster growing) strain. We find

\begin{equation}
x_{0}=l\frac{v_{1}^{2}}{v_{1}^{2}-v_{2}^{2}}=l\frac{(1+s)^{2}}{s(2+s)},
\label{eq:x0}
\end{equation}

\noindent where positive~$x_{0}$ corresponds to the direction towards the colony with the second strain, and negative~$x_{0}$ corresponds to the opposite direction (see figure~\ref{fig:collision_theory}). Similar to the selective sweep in the circular geometry, the less fit strain is eventually enclosed by the other strain. The time to this enclosure is given by

\begin{equation}
t_{f}=\frac{R_{b}+x_{0}-l}{v_{2}}=\frac{l}{v_{2}s},
\label{eq:collision_enclosure}
\end{equation}

\noindent as one can see from figure~\ref{fig:collision_theory} and equations~(\ref{eq:Rb}) and~(\ref{eq:x0}).

We derived sector and colony shapes using the equal-time argument and the results are consistent with our microscopic model. Because this argument relies only on the assumptions of constant expansion velocities and impenetrability of occupied regions, the equal-time argument applies more generally. As a result, we expect that the competition outcome is determined only by the ratio of expansion velocities of the two strains or species for any model consistent with these two assumptions. 

\section{Comparison with experiments}
\label{SComparison}

In the two preceding sections, we described theoretical predictions for the patterns of genetic diversity arising from spatial competitions. In section~\ref{SCSpatial}, we developed a generic reaction-diffusion model in order to predict macroscopic spatial patterns from microscopic parameters like cellular growth rates and effective diffusion constants. In section~\ref{SCAnalysis}, we then derived analytical formulae for boundaries between regions occupied by the competing strains using the equal-time argument.  
In this section, we experimentally test our theoretical predictions, using the expansion of the budding yeast \textit{Saccharomyces cerevisiae} on agar surfaces as a model system. We also employ the analytical formulae from the equal-time argument, equations~(\ref{EPhiET}), (\ref{ECircularSectorSolution}), (\ref{eq:Rb}), and~(\ref{eq:x0}), as a means to measure relative fitness~$s$ of two yeast strains. 

\begin{figure}
\includegraphics[width=8cm]{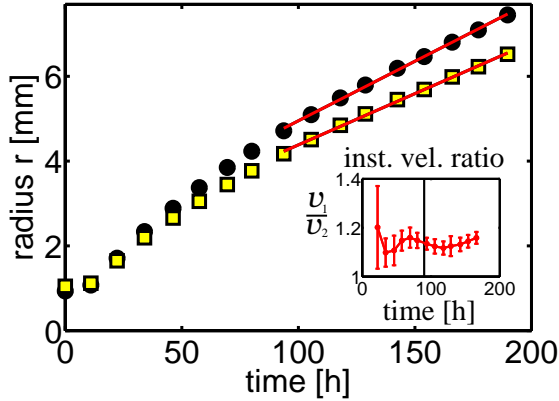}
\caption{(Colour online) The radii of yeast colonies as a function of time. Yellow squares and black circles correspond to colonies of the wild-type and the advantageous sterile mutant, respectively. After an initial transient (time $<$ 90h), described in more detail in the text, the radii are well fitted by a straight line (red), in accordance with a constant expansion velocity. 
Inset: Instantaneous velocity ratio as function of time. The instantaneous velocities at a specific time are determined from linear fits to the radii of the five surrounding time points. The black vertical line indicates the starting time for the fit in the main figure.}
\label{fig:velocities}
\end{figure} 

\subsection{Testing the equal-time argument}

We first tested the validity of the equal-time argument results of section~\ref{SCAnalysis} with a particular pair of \textit{S.\ cerevisiae} strains that have a large fitness difference: the wild-type and a faster growing mutant, which owes its advantage to the removal of a metabolically costly mating system; see the supplementary information~(section~S1) for more details. 
%
The equal-time argument assumes spatial expansion at a constant velocity. To see whether this assumption was valid in our experimental system, we measured the increase of the radius of circular yeast colonies over time, see figure~\ref{fig:velocities}. During the first day, colonies barely grow, presumably because it takes time for the populations to reach the carrying capacity in the spatial region of the inoculum. Afterwards, the colony fronts expand at about $20\mu$m/h. The rate of expansion first slows down gradually over time, presumably because of nutrient depletion and drying out of the agar gel. For larger times, a stationary expansion front has established, and colonies grow at a constant expansion velocity. Since the latter is a pre-requisite for the equal-time argument to apply, we only consider these later times in our further analysis. A similar behaviour was observed for the expansion of yeast colonies from a linear inoculation (data not shown), although the absolute values of the velocities were very different; e.g. for the wild-type, we obtained $v_{2}=14.5\pm1.9\;\mu$m/h for linear vs. $v_{2}=25.0\pm0.5\;\mu$m/h for circular expansions. This difference is presumably due to larger colonies and therefore more severe nutrient depletion in linear inoculations. 

\begin{figure}
(a) \includegraphics[width=3.5cm]{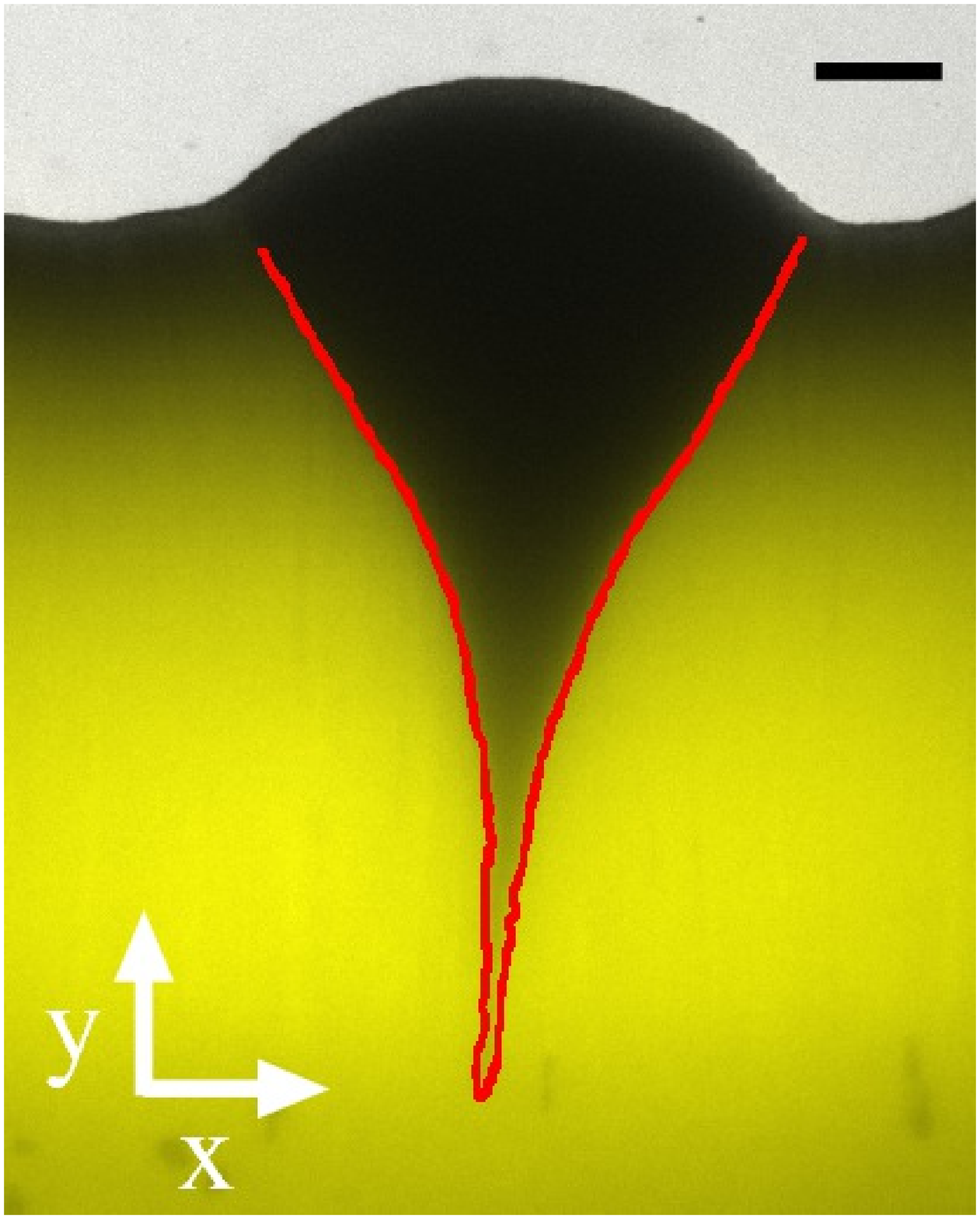}\hspace*{0.5em}
(b) \includegraphics[width=5.5cm]{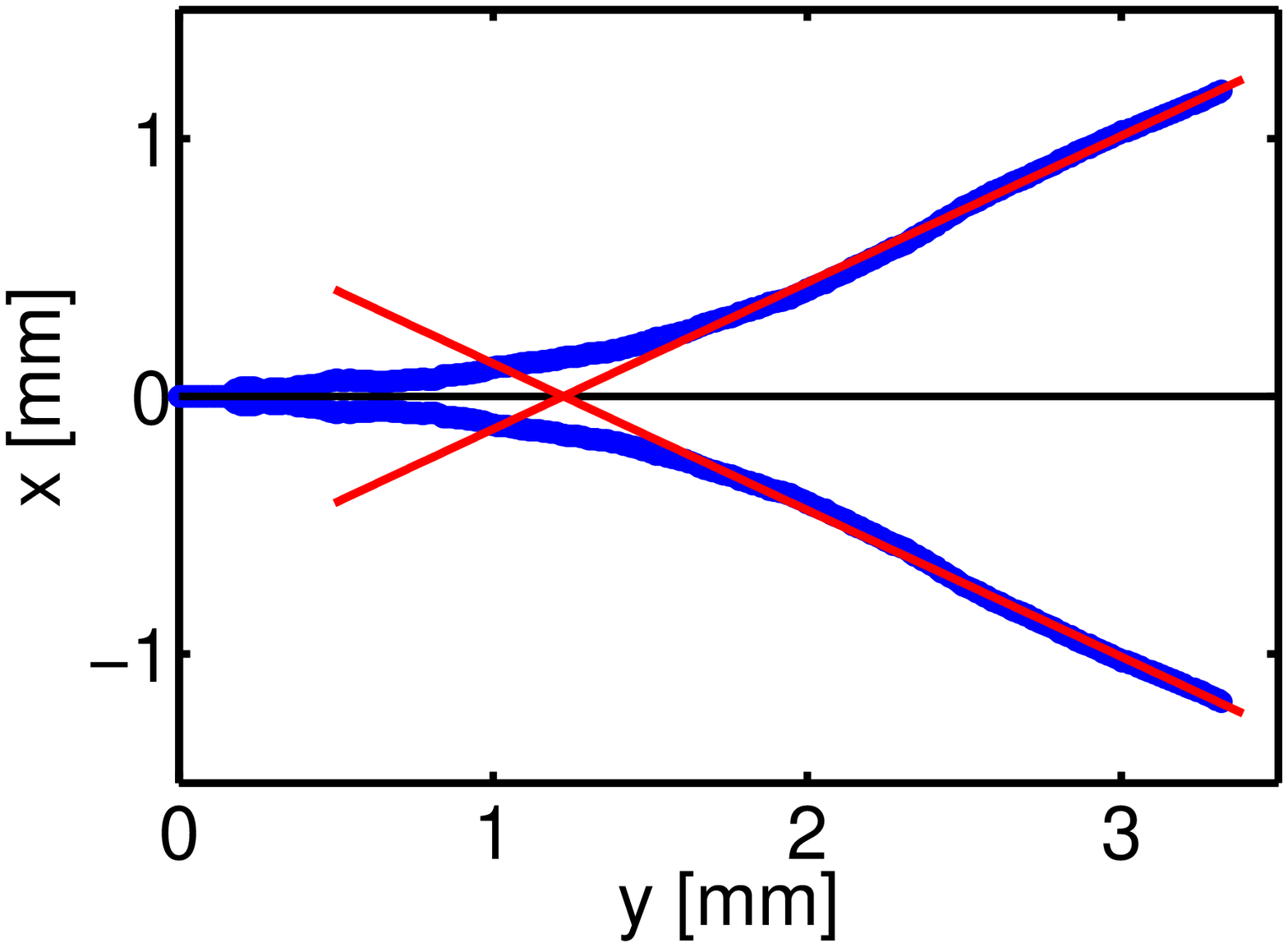} \\
\caption{(Colour online) Fitness estimation from linear expansion sectors. 
(a) \textit{S.~cerevisiae} colony grown from a linear inoculation at the bottom of the picture. A sector of the advantageous sterile mutant~(black) emerges in the predominantly wild-type colony~(yellow). The sector boundaries inferred from the image are shown with red lines. The scale bar is 500$\;\mu$m. 
(b) Sector boundaries~(blue dots) extracted from the image shown in (a), and fits~(lines, $r^2>0.995$) to equation~(\protect{\ref{EPhiET}}). Note that the early part of sector growth differs from the later part. One possible explanation for this difference is the sector establishment process discussed in Sec.~\protect{\ref{SCSpatial}} when sector boundaries are not fully separated and interact with each other. The equal-time argument does not apply in this case.}
\label{fig:linear_sectors}
\end{figure}

\begin{figure}
(a) \includegraphics[width=7cm]{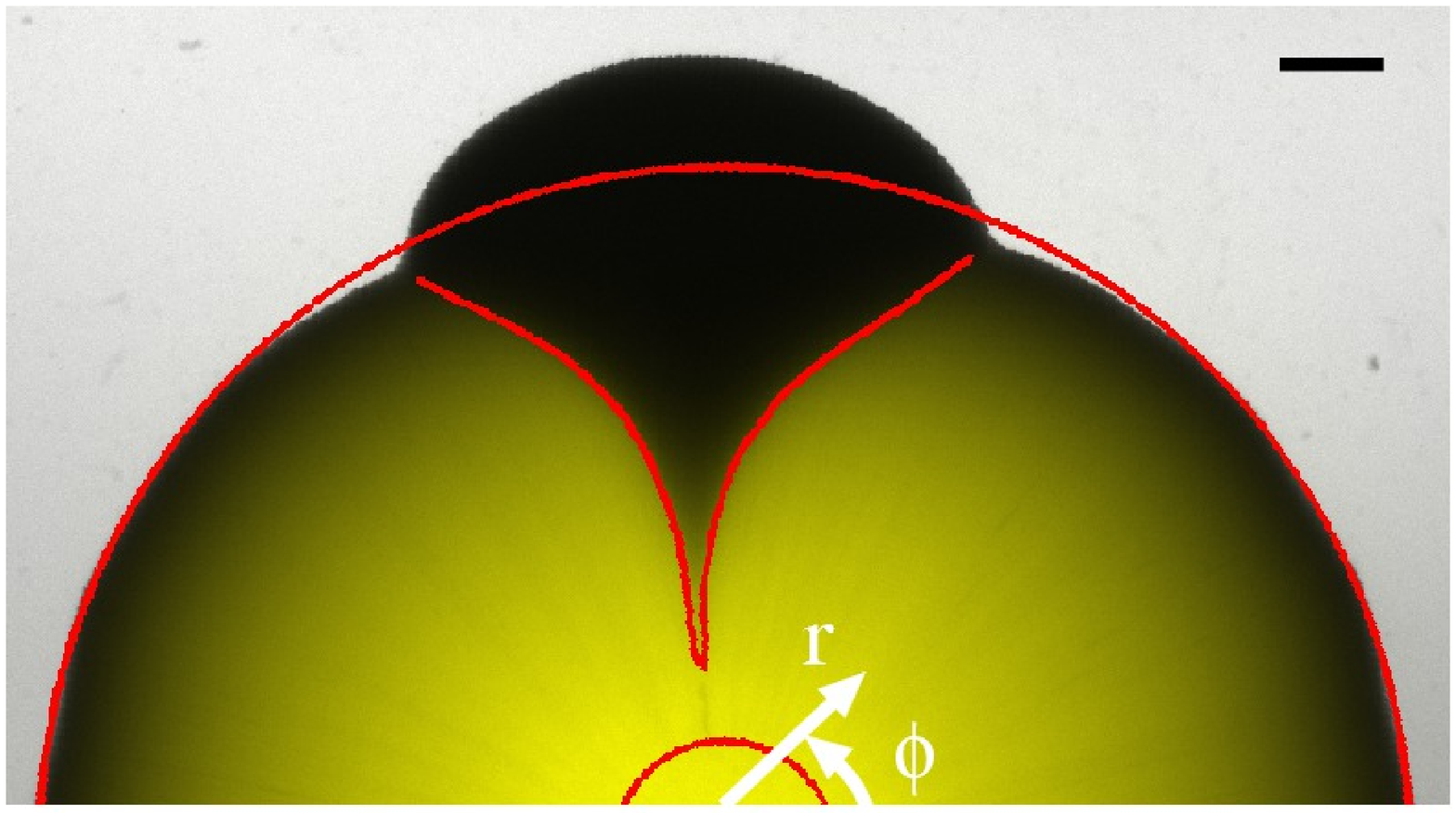}\hspace*{0.5em}
(b) \includegraphics[width=6cm]{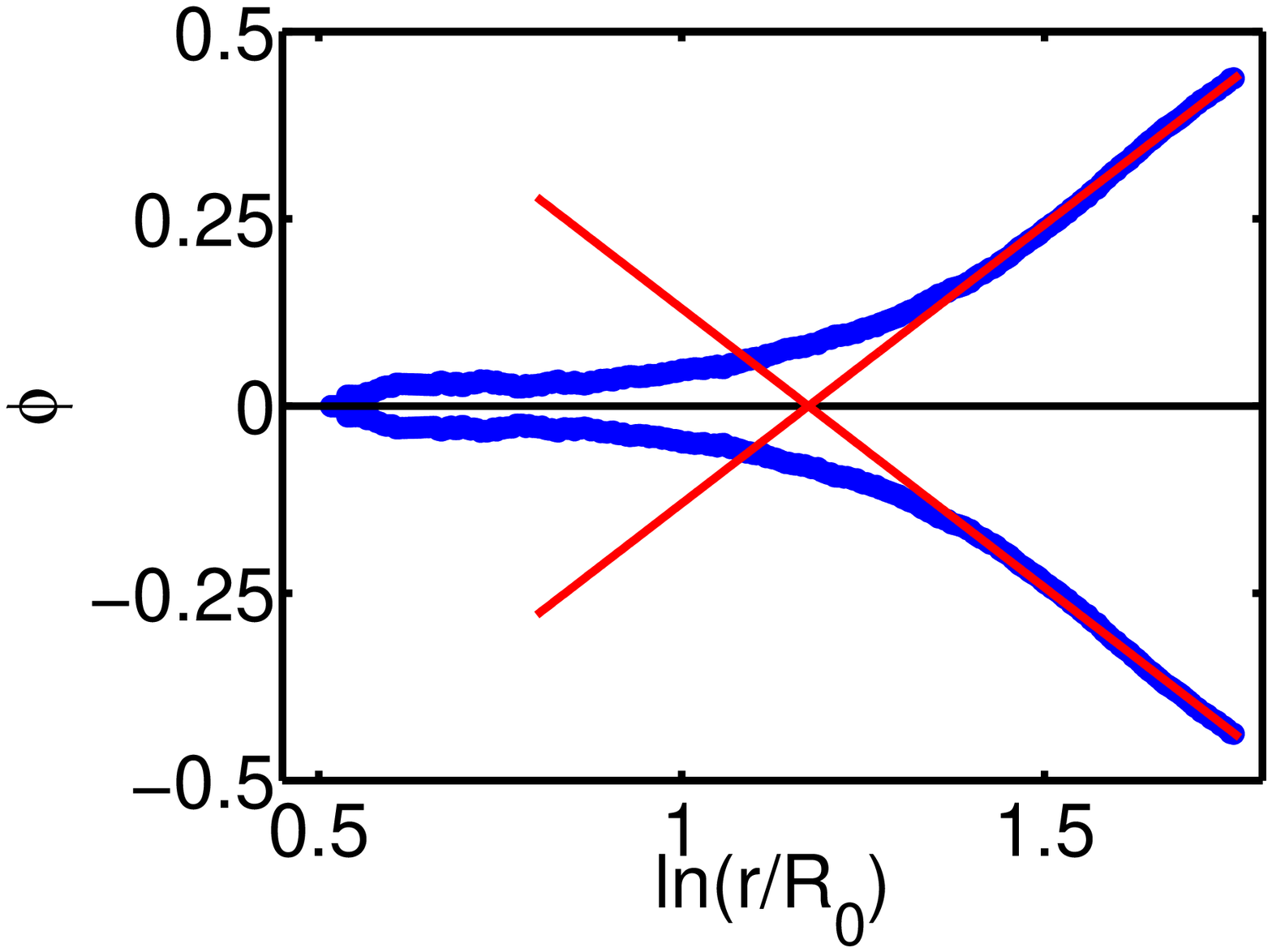} 
\caption{(Colour online) Fitness estimation from radial expansion sectors. The same as figure~\protect{\ref{fig:linear_sectors}}, but for a circular geometry. In (a), only the top half of a circular colony is shown. The smaller red circle shows the inoculum, and the larger red circle marks the colony radius. The scale bar is~$1\;$mm.}
\label{fig:circular_sectors}
\end{figure}

\begin{figure}
\begin{tabular}{l}
\includegraphics[width=4.5cm]{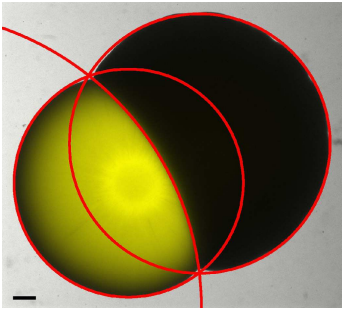} \\
\end{tabular}
\caption{(Colour online) Fitness estimation from colony collisions. The wild-type~(yellow) colony meets the colony of the advantageous sterile mutant~(black). The red lines are the fits of colony boundaries by circles. The relative fitness of the colonies can be measured from the radius and center of the circle fitted to the interface between the colonies; see equations~(\protect{\ref{eq:Rb}}) and~(\protect{\ref{eq:x0}}). The scale bar is~$1\;$mm.}
\label{fig:collisions}
\end{figure}

We next tested whether the equal-time argument correctly predicts shapes of boundaries between the strains: linear for linear expansions, logarithmic spirals for circular expansions, and circular for colliding colonies. This was indeed the case, see figures~\ref{fig:linear_sectors},~\ref{fig:circular_sectors}, and~\ref{fig:collisions}. Note that in the case of sectors, the equal-time argument applies only at later times. During earlier times, the expansion velocity is not constant, as discussed above. In addition, during this initial transient, sectors have not yet fully established, see the discussion in section~\protect{\ref{SCSpatial}} and in the supplement section~S5. Since both constant expansion velocity and fully established domains of each strains are required for the equal-time argument, we fit our theoretical predictions equations~(\ref{EPhiET}) and~(\ref{ECircularSectorSolution}) to the sector shapes only for later times, finding excellent agreement ($r^2>0.995$).

\subsection{Measuring relative fitness}

This agreement allows us to use the  analytical results from the equal-time argument in order to measure the relative fitness $s$. From our analysis, the relative fitness can be estimated by five different methods, using (i-ii) ratio of expansion velocities~$v_{1}$ and~$v_{2}$ of isolated colonies in linear and circular geometries, (iii-iv) sector shapes in linear and circular geometries using equations~(\ref{EPhiET}) and (\ref{ECircularSectorSolution}), and (v) the interface shape of colliding circular colonies using equations (\ref{eq:Rb},\ref{eq:x0}). For consistency, we used the same time window for all assays, see the supplementary information~(section~S1) for details. To provide a reference, we also measured the relative growth rate during exponential phase in a well-mixed test tube, either in separate or in mixed cultures. 

                                                                                                                                                                                                                                                                                                                                                                                                                                                                                                                                                                                                                                                                                                                                                                                                                                                                                                                                                                                               \begin{table}
\begin{tabular}{l@{\;\;}|l@{\;\;}|l}\hline\hline
Assay							 & Method    	      & Selective advantage,$\;s$ \\\hline
\multirow{2}{*}{Linear expansion}& velocity ratios	  & $0.10\pm 0.16\;\; (N=11)$ \\
								 & sectors 			  & $0.20\pm 0.13\;\; (N=23)$ \\\hline
\multirow{3}{*}{Radial expansion}& velocity ratios	  & $0.16\pm 0.08\;\; (N=19)$ \\
								 & sectors	 		  & $0.23\pm 0.04\;\; (N=24)$ \\
								 & colony collisions  & $0.17\pm 0.02\;\; (N=9)$\\\hline
\multirow{2}{*}{Liquid culture}  & growth rate ratios & $0.17\pm 0.03\;\; (N=3)$ \\
								 & competitions		  & $0.18\pm 0.02\;\; (N=3)$ \\
\hline\hline
\end{tabular}
\caption{Comparison of relative fitnesses measured by different methods. Errors are standard deviations (not standard errors of the mean). The number of replicates~($N$) is given in parentheses. Note that the accuracy of different assays varies by about an order of magnitude. The large standard deviations for linear expansions are, at least partially, due to front undulations, which make sector boundaries irregular and sector angles more variable. Sector and collision measurement have smaller standard deviations compared to direct velocity measurements. We attribute this distinction to the fact that both strains experience exactly the same local environment in the sector assay, but only approximately the same environment in the velocity assay. In particular, some environmental parameters--like the local dryness of the agar gel--are hard to control, and even identically prepared Petri dishes inevitably have slight differences in these parameters. Such variations affect velocity measurements, where the two strains are grown on two different (but identically prepared) Petri dishes, but do not affect sector measurements, where the strains are grown on the same Petri dish, and compete at the same point in space.}
\label{tab:fitnesses}
\end{table} 

The relative fitnesses $s$ obtained from the different fitness assays yield quite similar results, summarized in table~\ref{tab:fitnesses}. Indeed, the measurements are not significantly different from each other, except for the circular sector result which differs significantly from both the liquid culture competition and the colony collision assays ($p<0.05$, see the supplement section~S6 for details on the statistical testing procedure). The deviation for circular sectors could be caused by a systematic error in sector analysis. 
However, a likely explanation of the disagreement between different fitness estimates is some additional spatial structure not accounted for in our theory. Indeed, yeast colonies do not only expand on the surface of a Petri dish, but they also thicken over time to a height of about~$1\;$mm, which is neglected in our two-dimensional theory. It is therefore possible that the advantageous mutant grows on top of the wild-type, producing an apparently larger sector and leading to an overestimate of~$v_{1}/v_{2}$.

Table~\ref{tab:fitnesses} shows that the different fitness assays have standard deviations that vary over an order of magnitude, and therefore have very different accuracies.
%
Expansion velocities of isolated single-strain colonies are the most straightforward measurement of fitness on a plate. However, they are less accurate than the sector and colony collision assays, as reflected by their high standard deviations in table~\ref{tab:fitnesses} and the fluctuations of the instantaneous velocity ratio in the inset of figure~\ref{fig:velocities}. In sector and collision assays, the two competing strains are in exactly the same environment, and inevitable slight differences in the experimental conditions, such as humidity of the agar, influence both strains equally. This is not true for the isolated colonies of the expansion velocity assays, which are therefore more variable. Even more important, we found that relative fitnesses obtained from experiments on different batches of plates were significantly different when determined from expansion velocities, but not when determined from sectors or colony collisions; see the supplementary information~(section~S1). Nevertheless, the ratio of the expansion velocities is similar for linear and circular expansions, see table~\ref{tab:fitnesses}, despite a large difference in the absolute values of the velocities, as discussed above.

Linear expansion assays also have large standard deviations. This is probably due to undulations of linear fronts~\cite{Kessler:Instabilities}, clearly visible in figure~\ref{FNSComparison}, which distort the sector shapes. These undulations are significantly reduced in the circular geometry, leading to smaller standard deviations. It is thus advantageous to determine fitness from circular rather than linear expansions.

From this discussion and table~\ref{tab:fitnesses}, it follows that the radial expansion sector and colony collision assays are the most reliable assays to measure relative fitness in a spatial competition. We therefore only use these two spatial fitness assays in the following.

\begin{figure}
\begin{tabular}{ll}
\includegraphics[width=7cm]{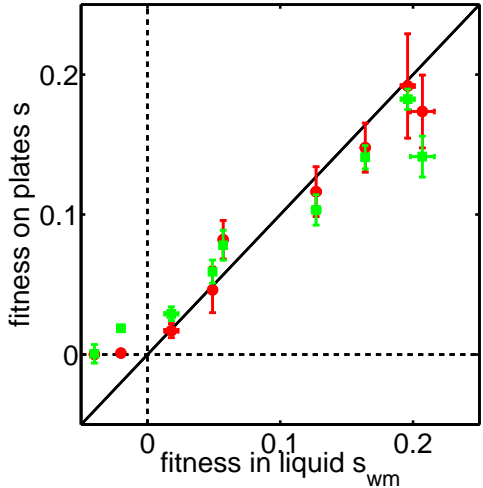} \\
\end{tabular}
\caption{(Colour online) Comparison of the selective advantage in well-mixed liquid culture and in spatial expansions. We varied the relative fitness using the drug cycloheximide for competitions of the cycloheximide-sensitive wild-type with a cycloheximide-resistant mutant. The fitness $s$ measured with radial expansion sectors (red circles) and colony collisions (green squares) agrees well with the fitness $s_{\rm wm}$ from the liquid competition assay, since all points lie close to the diagonal (black line, not a fit). The agreement of the spatial fitness $s$ with the liquid fitness $s_{\rm wm}$ is predicted by our theoretical model when migration is driven by cell growth, see the corresponding figure~\ref{fig:fitness_plate_liquid}.}
\label{fig:cycloheximide}
\end{figure}

We have so far found agreement between experiments and our equal-time argument predictions, but only for one pair of strains with one particular fitness difference. Therefore, we performed further experiments over a range of fitness values. To this purpose, we competed a strain resistant to with a strain sensitive to cycloheximide (a drug inhibiting translation) for varying cycloheximide concentrations in the medium. The relative fitness of the resistant strain, as e.g., measured with a liquid culture competition assay, increases linearly with the drug concentration, see supplementary figure~S8. For all concentrations tested, radial sectors and colony collision boundaries could be well fitted with logarithmic spirals and circles, respectively, as predicted by our theory.
More importantly, we were able to compare the fitness advantage in liquid culture~$s_{\rm{wm}}$ to the fitness advantage on Petri dishes~$s$ over a wide range of relative fitnesses. We found good agreement between the two spatial assays and the liquid competition assay, see figure~\ref{fig:cycloheximide}. 

\subsection{Testing the reaction-diffusion model}

Our reaction-diffusion model predicts an agreement of the spatial and well-mixed relative fitness, if migration is driven by cell growth, i.e.~$D_{01}\sim g_{1}$ and~$D_{02}\sim g_{2}$, as discussed at the end of section~\ref{SCSpatial}. The results shown in figure~\ref{fig:cycloheximide} are therefore a direct confirmation of this version of the reaction-diffusion model. In addition, the reaction-diffusion model predicts a constant front expansion velocity, given by equation~\ref{EFisherVelocityFull}, which is indeed observed in our experiments for large times, as shown in figure~\ref{fig:velocities}. Furthermore, the reaction-diffusion model gives rise to the same, experimentally confirmed, macroscopic spatial pattern predicted by the equal-time argument, independent of microscopic details on the cellular length scale. We therefore preformed competition experiments with \textit{S.~ cerevisiae} strains that have different cell division patterns, as well as with the bacterium \textit{Pseudomonas aeruginosa}, see supplementary information (section~S3). All experiments could be well described by our theory.

In summary, there are experimental subtleties that our phenomenological theory does not take into account, such as the expansion velocity slowdown, yeast colony thickness, or possible strain interactions~\footnote{The mutualistic or antagonistic interaction represented by the terms with~$\epsilon_{1}$ and~$\epsilon_{2}$ in equation~(\ref{ECompetitionSpatialFull}) could also change the relative fitness in experiments where the strains are in physical contact compared to experiments where the strains are grown in isolation. Three dimensional yeast colonies have a relatively large contact angle with the agar surface at the colony edge~\cite{nguyen:instability}. Therefore, the effect of such hypothetical interactions between the two strains might not be negligible if the density of yeast cells at the colony edge is not sufficiently small, as it would be if cell density decayed exponentially at the frontier.}. All these effects could contribute to the slight differences of the fitness values determined by different methods, see table~\ref{tab:fitnesses} and figure~\ref{fig:cycloheximide}. Nevertheless, our theory describes the shapes of established sectors and colony collisions very accurately. Considering that the methods to determine relative fitness are very different, it is remarkable that the obtained fitness values are so similar, in particular results from well-mixed liquid culture and spatial growth on agar surfaces.

\section{Discussion}
\label{SConclusions}

Natural selection in well-mixed populations leads to selective sweeps of beneficial genotypes occurring exponentially fast in time. However, in spatially expanding populations, competition results in more complicated temporal and spatial patterns. Since both the advantageous and deleterious genotypes can spread into uncolonized territories, their competition can result in sectoring patterns like that shown in figure~\ref{FSCICircularSolution}. Sectoring spatial patterns provide an alternative fitness assay to the commonly used assays based on competition in a well-mixed environment of a test tube. This alternative facilitates spatial evolutionary experiments, which might contribute to understanding adaptations in different environments.

The spatial assay may also be more accurate, provided front undulations, nutrient depletion, variations in agar wetness, and other experimental complications can be overcome. For small fitness differences, the assay could in principle acquire sensitivity because it measures~$\sqrt{v_{1}^{2}/v_{2}^{2}-1}=\sqrt{s(2+s)}\approx\sqrt{2s}$ instead of~$v_{1}/v_{2}=1+s$~\cite{Hallatschek:LifeFront}. For large fitness differences, higher accuracy could also result from longer observation times, as the deleterious strain survives longer in spatial settings. A deleterious strain is eliminated linearly~(linear geometry) or logarithmically~(circular geometry) in time, unlike in the well-mixed environment, where it is eliminated exponentially fast. More important, a spatial competition experiment could be superior to a well-mixed one when used for screening for beneficial mutations. On a Petri dish, many beneficial mutations can be assayed in parallel from expansions started by a small fraction of fluorescently-labeled mutagenized cells mixed with wild-type cells. In addition, each mutation is spatially isolated, and a dense aggregate of cells only a few generations away from the original mutation could be easily collected for future use.

Our analysis of competition during range expansions has applications for evolutionary dynamics in spatially extended habitats as well. In particular, the predictions for one of the most important quantities in evolutionary dynamics, the duration of a selective sweep, is substantially different between spatial and nonspatial models. Deleterious genotypes persist much longer in spatial populations because the beneficial mutations spreading by Fisher waves may have to travel large distances needed, e.g., to engulf the wild-type populations, as in figure~\ref{FHeart}.

\ack{
DRN and AWM acknowledge conversations with P. Hersen. Overall support for this project was provided by the National Science Foundation, through Grant DMR-1005289, National Institute of General Medical Sciences Grant P50GM068763 of the National Centers for Systems Biology, and by the Harvard Materials Research Science and Engineering Center through DMR-0820484.
}

\section*{References}

\bibliography{SCBibs}

\begin{thebibliography}{10}

\bibitem{Darwin:Origin}
C.~Darwin.
\newblock {\em {On the origin of species by means of natural selection, or the
  preservation of favoured races in the struggle for life}}.
\newblock New York University Press, New York, 1859.

\bibitem{Barton:Evolution}
N.~H. Barton, D.~E.~G. Briggs, J.~A. Eisen, D.~B. Goldstein, and N.~H. Patel.
\newblock {\em {Evolution}}.
\newblock Cold Spring Harbor Laboratory Press, Cold Spring Harbor, 2007.

\bibitem{Elena:EvolutionReview}
S.~F. Elena and R.~E. Lenski.
\newblock {Evolution experiments with microorganisms: the dynamics and genetic
  bases of adaptation}.
\newblock {\em Nature Reviews: Genetics}, 4:457, 2003.

\bibitem{Ron:Sectors}
I.~G. Ron, I.~Golding, B.~Lifsitz-Mercer, and E.~Ben-Jacob.
\newblock {Bursts of sectors in expanding bacterial colonies as a possible
  model for tumor growth and metastases}.
\newblock {\em Physica A}, 320:485, 2003.

\bibitem{lambert:analogy}
G.~Lambert, L.~Estvez-Salmeron, S.~Oh, D.~Liao, B.~M. Emerson, T.~D. Tlsty, and
  R.~H. Austin.
\newblock {An analogy between the evolution of drug resistance in bacterial
  communities and malignant tissues}.
\newblock {\em Nature Reviews Cancer}, 11:375--382, 2011.

\bibitem{Korolev:Review}
K.~S. Korolev, M.~Avlund, O.~Hallatschek, and D.~R. Nelson.
\newblock {Genetic demixing and evolution in linear stepping stone models}.
\newblock {\em Reviews of Modern Physics}, 82:1691--1718, 2010.

\bibitem{kerr:tragedy}
B.~Kerr, C.~Neuhauser, B.~J.~M. Bohannan, and A.~M. Dean.
\newblock {Local migration promotes competitive restraint in a host--pathogen
  'tragedy of the commons'}.
\newblock {\em Nature}, 442:75--78, 2006.

\bibitem{Coberly:SpaceCoEvolution}
L.~C. Coberly, W.~Wei, K.~Y. Sampson, J.~Millstein, H.~A. Wichman, and S.~M.
  Krone.
\newblock {Space, Time, and Host Evolution Facilitate Coexistence of Competing
  Bacteriophages: Theory and Experiment}.
\newblock {\em The American Naturalist}, 173:121, 2009.

\bibitem{hauert:snowdrift}
C.~Hauert and M.~Doebeli.
\newblock {Spatial structure often inhibits the evolution of cooperation in the
  snowdrift game}.
\newblock {\em Nature}, 428:643--646, 2004.

\bibitem{rainey:radiation}
P.~B. Rainey and M.~Travisano.
\newblock {Adaptive radiation in a heterogeneous environment}.
\newblock {\em Philosophical Transactions of the Royal Society B}, 280:29--101,
  1977.

\bibitem{wakita:expansion}
J.~Wakita, K.~Komatsu, A.~Nakahara, T.~Matsuyama, and M.~Matsushita.
\newblock {Experimental Investigation on the Validity of Population Dynamics
  Approach to Bacterial Colony Formation}.
\newblock {\em Journal of the Physical Society of Japan}, 63:1205--1211, 1994.

\bibitem{benjacob:cooperative_growth}
E.~Ben-Jacob, O.~Schochet, A.~Tenenbaum, I.~Cohen, A.~Czir{\'o}k, and
  T.~Vicsek.
\newblock Generic modelling of cooperative growth patterns in bacterial
  colonies.
\newblock {\em Nature}, 368:46--49, 1994.

\bibitem{benjacob:complex_patterns}
E.~Ben-Jacob, I.~Cohen, O.~Shochet, I.~Aranson, H.~Levine, and L.~Tsimring.
\newblock Complex bacterial patterns.
\newblock {\em Nature}, 373:566--567, 1995.

\bibitem{golding:sectors}
I.~Golding, I.~Cohen, and E.~Ben-Jacob.
\newblock Studies of sector formation in expanding bacterial colonies.
\newblock {\em Europhysics Letters}, 48:587, 1999.

\bibitem{HallatschekNelson:ExperimentalSegregation}
O.~Hallatschek, P.~Hersen, S.~Ramanathan, and {D. R. Nelson}.
\newblock {Genetic drift at expanding frontiers promotes gene segregation}.
\newblock {\em Proc. Natl. Acad. Sci. USA}, 104:19926, 2007.

\bibitem{beer:deadly}
A.~Be'Er, H.~P. Zhang, E.~L. Florin, S.~M. Payne, E.~Ben-Jacob, and H.~L.
  Swinney.
\newblock Deadly competition between sibling bacterial colonies.
\newblock {\em Proceedings of the National Academy of Sciences}, 106:428, 2009.

\bibitem{korolev:neutral_expansions}
K.~S. Korolev, J.~B. Xavier, D.~R. Nelson, and K.~R. Foster.
\newblock {A Quantitative Test of Population Genetics Using Spatio-Genetic
  Patterns in Bacterial Colonies}.
\newblock {\em The American Naturalist}, 178:538, 2011.

\bibitem{KerrBohannan02}
B.~Kerr, M.~A. Riley, M.~W. Feldman, and B.~J.~M. Bohannan.
\newblock Local dispersal promotes biodiversity in a real-life game of
  rock-paper-scissors.
\newblock {\em Nature}, 418:171--174, 2002.

\bibitem{HabetsDeVisser06}
M.~G. J.~L. Habets, D.~E. Rozen, R.~F. Hoekstra, and J.~A. G.~M. de~Visser.
\newblock The effect of population structure on the adaptive radiation of
  microbial populations evolving in spatially structured environments.
\newblock {\em Ecological Letters}, 9:1041--1048, 2006.

\bibitem{Harcombe10}
W.~Harcombe.
\newblock Novel cooperation experimentally evolved between species.
\newblock {\em Evolution}, 64:2166--2172, 2010.

\bibitem{SaxerTravisano09}
G.~Saxer, M.~Doebeli, and M.~Travisano.
\newblock Spatial structure leads to ecological breakdown and loss of
  diversity.
\newblock {\em Proceedings of the Royal Society B: Biological Sciences},
  276:2065--2070, 2009.

\bibitem{VelicerYu03}
G.~J. Velicer and Y.~T.~N. Yu.
\newblock Evolution of novel cooperative swarming in the bacterium myxococcus
  xanthus.
\newblock {\em Nature}, 425:75--78, 2003.

\bibitem{benjacob:review}
E.~Ben-Jacob, I.~Cohen, and H.~Levine.
\newblock Cooperative self-organization of microorganisms.
\newblock {\em Advances in Physics}, 49:395--554, 2000.

\bibitem{Hallatschek:LifeFront}
O.~Hallatschek and D.~R. Nelson.
\newblock {Life at the front of an expanding population}.
\newblock {\em Evolution}, 64:193, 2010.

\bibitem{hallatschek:noisy_wave}
O.~Hallatschek.
\newblock The noisy edge of traveling waves.
\newblock {\em Proceedings of the National Academy of Sciences USA},
  108:1783--1787, 2011.

\bibitem{Kessler:Instabilities}
D.~A. Kessler and H.~Levine.
\newblock {Fluctuation-induced diffusive instabilities}.
\newblock {\em Nature}, 394:556, 1998.

\bibitem{pipe:colonies}
L.~Z. Pipe and M.~J. Grimson.
\newblock Spatial-temporal modelling of bacterial colony growth on solid media.
\newblock {\em Mol. BioSyst.}, 4:192--198, 2008.

\bibitem{nadell:structure_plos}
C.~D. Nadell, K.~R. Foster, and J.~B. Xavier.
\newblock Emergence of spatial structure in cell groups and the evolution of
  cooperation.
\newblock {\em PLoS Computational Biology}, 6:e1000716, 2010.

\bibitem{Murray:MathematicalBiology}
J.~D. Murray.
\newblock {\em {Mathematical Biology}}.
\newblock Springer, 2003.

\bibitem{pearl:logistic}
R.~Pearl.
\newblock {The growth of populations}.
\newblock {\em The Quarterly Review of Biology}, 2:532--548, 1927.

\bibitem{carlson:logistic}
T.~Carlson.
\newblock {\"U}ber geschwindigkeit und gr{\"o}sse der hefevermehrung in
  w{\"u}rze.
\newblock {\em Biochem. Z}, 57:313--334, 1913.

\bibitem{nguyen:instability}
B.~Nguyen, A.~Upadhyaya, A.~Van~Oudenaarden, and M.~P. Brenner.
\newblock Elastic instability in growing yeast colonies.
\newblock {\em Biophysical Journal}, 86:2740--2747, 2004.

\bibitem{Fisher:FisherWave}
R.~A. Fisher.
\newblock {The wave of advance of advantageous genes}.
\newblock {\em The Annals of Eugenics}, 7:353, 1937.

\bibitem{Kolmogorov:FKPPEquation}
A.~N. Kolmogorov, N.~Petrovsky, and N.~S. Piscounov.
\newblock A study of the equation of diffusion with increase in the quantity of
  matter, and its application to a biological problem.
\newblock {\em Moscow University Bulletin of Mathematics}, 1:1, 1937.

\bibitem{Hallatschek:FisherWave}
O.~Hallatschek and {K. S. Korolev}.
\newblock {Fisher Waves in the Strong Noise Limit}.
\newblock {\em Physical Review Letters}, 103:108103, 2009.

\bibitem{Born:Optics}
M.~Born, E.~Wolf, and A.~B. Bhatia.
\newblock {\em {Principles of Optics}}.
\newblock Pergamon Press, 1964.

\bibitem{cook:curves_life}
T.~A. Cook.
\newblock {\em {The Curves of Life}}.
\newblock Dover, London, 1985.

\end{thebibliography}

\bibliographystyle{unsrt}

\end{document}